\documentclass{article}

 \usepackage[preprint]{neurips_2026}

\usepackage[utf8]{inputenc} 
\usepackage[T1]{fontenc}    
\usepackage{hyperref}       
\usepackage{url}            
\usepackage{booktabs}       
\usepackage{amsfonts}       
\usepackage{nicefrac}       
\usepackage{microtype}      
\usepackage{xcolor}         

\usepackage{placeins}
\usepackage{dblfloatfix}

\usepackage{microtype}
\usepackage{hyperref}
\usepackage{url}
\usepackage{booktabs}
\usepackage{graphicx}
\usepackage{amsmath}
\usepackage{amsthm}

\usepackage{multirow}
\usepackage{makecell}
\usepackage{algorithm}
\usepackage{algpseudocode}
\usepackage{pifont}
\usepackage{wrapfig}
\usepackage{tocloft}
\usepackage{etoc} 

\usepackage{etoc}
\usepackage{caption}

\usepackage{booktabs}
\usepackage{multirow}
\usepackage{makecell}
\usepackage{array}
\usepackage[edges]{forest}
\usetikzlibrary{shadows}

\usepackage{tikz}
\usetikzlibrary{arrows.meta,positioning}
\usepackage{array}
\usepackage[table]{xcolor}
\newcommand{\legenditem}[2]{%
  \textcolor{#1}{\rule{0.9em}{0.9em}}\hspace{0.3em}\textit{#2}%
}
\usepackage{caption}
\definecolor{pastelblue}{RGB}{180,210,235}
\definecolor{pastelgreen}{RGB}{190,225,200}
\definecolor{pastelpink}{RGB}{235,190,205}
\definecolor{pastelyellow}{RGB}{245,225,170}
\definecolor{pastelpurple}{RGB}{210,200,235}
\definecolor{linegray}{RGB}{95,95,95}
\definecolor{gridgray}{RGB}{150,150,150}
\definecolor{headergray}{RGB}{245,246,248}
\definecolor{schematicbg}{RGB}{250,250,251}

\usepackage{tikz}
\usetikzlibrary{positioning,arrows.meta}

\definecolor{pastelblue}{RGB}{180,210,235}
\definecolor{pastelgreen}{RGB}{190,225,200}
\definecolor{pastelyellow}{RGB}{245,225,170}
\definecolor{pastelpink}{RGB}{235,190,205}
\definecolor{pastelpurple}{RGB}{210,200,235}
\definecolor{linegray}{RGB}{95,95,95}

\tikzset{
  evnode/.style={
    circle,
    draw=linegray,
    line width=0.8pt,
    minimum size=5.2mm,
    inner sep=0pt
  },
  evline/.style={
    draw=linegray,
    line width=0.75pt,
    -{Latex[length=1.4mm,width=1.1mm]}
  }
}

\newcolumntype{C}[1]{>{\centering\arraybackslash}p{#1}}

\usepackage{lineno}

\definecolor{darkblue}{rgb}{0, 0, 0.5}
\hypersetup{colorlinks=true, citecolor=darkblue, linkcolor=darkblue, urlcolor=darkblue}

\title{Do Agent Societies Develop Intellectual Elites? \\The Hidden Power Laws of Collective Cognition in LLM Multi-Agent Systems}

\author{%
  Kavana Venkatesh \\
  Department of Computer Science\\
  Virginia Tech\\
  Blacksburg, VA \\
  \texttt{kavanav@vt.edu} \\
  \And
  Jiaming Cui \\
  Department of Computer Science \\
  Virginia Tech\\
  Blacksburg, VA \\
  \texttt{jiamingcui@vt.edu} 
}

\begin{document}

\maketitle

\begin{abstract}
Large Language Model (LLM) multi-agent systems are increasingly deployed as interacting agent societies, yet scaling these systems often yields diminishing or unstable returns, the causes of which remain poorly understood. We present the first large-scale empirical study of coordination dynamics in LLM-based multi-agent systems, introducing an atomic event-level formulation that reconstructs reasoning as cascades of coordination. Analyzing over 1.5 Million interactions across tasks, topologies, and scales, we uncover three coupled laws: coordination follows heavy-tailed cascades, concentrates via preferential attachment into intellectual elites, and produces increasingly frequent extreme events as system size grows. We show that these effects are coupled through a single structural mechanism: an integration bottleneck, in which coordination expansion scales with system size while consolidation does not, producing large but weakly integrated reasoning processes. To test this mechanism, we introduce Deficit-Triggered Integration (DTI), which selectively increases integration under imbalance. DTI improves performance precisely where coordination fails, without suppressing large-scale reasoning. Together, our results establish quantitative laws of collective cognition and identify coordination structure as a fundamental, previously unmeasured axis for understanding and improving scalable multi-agent intelligence.
\end{abstract}
\section{Introduction}
\label{sec:intro}

LLM multi-agent systems (MAS) have been widely used in planning, coding, and deliberative reasoning tasks~\cite{chen2023agentverse, qian2024chatdev, du2024improving, venkatesh2026physicsagentabm}. However, a major challenge lies in their reliability under scaling: adding more agents does not naturally yield proportional gains; instead, performance may plateau, oscillate, or even degrade beyond a certain tipping point~\cite{chen2024more,cemri2025multi}. Despite substantial efforts in prompting and reasoning~\cite{wei2022chain,yao2022react}, these failures remain stubborn, indicating that the problem is structural and systematic.

These dynamics run deeper than current benchmarks capture. Existing metrics for LLM MAS: task success, final-answer accuracy, and cumulative reward~\cite{liu2023agentbench}, are largely centered on outcomes while overlooking the internal dynamics of collective reasoning. As a result, they cannot distinguish coordinated synthesis from fragmented processes that coincidentally reach correct answers~\cite{zhuge2024gptswarm, cemri2025multi}, nor capture how effort is distributed, influence accumulates, or coordination stabilizes. Consequently, while prior analyses document scaling failures, they do not explain them systematically~\cite{chen2024more,cemri2025multi}, leaving architectural choices heuristic and interventions reactive~\cite{guo2024large}. 

What is missing is a science of coordination dynamics; not what agents produce, but how reasoning is initiated, propagated, contested, and resolved across them~\cite{guo2024large}. Therefore, in this paper, we show that LLM multi-agent systems exhibit cascade-driven dynamics, in which agents interact iteratively and respond to prior outputs~\cite{du2024improving}, while delegation and contradiction induce branching cascades. The final output is thus shaped by the full tree of downstream events that an initial reasoning step generates~\cite{liang2024encouraging}. These interaction patterns reveal that coordination is not evenly distributed, but rather organized through expanding and competing cascades of reasoning activity: claims that attract early engagement are more likely to recruit further coordination, suggesting a reinforcement effect in how reasoning activity is routed and amplified. Meanwhile, the unfolding of such coordination is constrained by factors such as limited context~\cite{packer2310memgpt}, communication bandwidth, and token budgets~\cite{gu2024your}. Taken together, these characteristics suggest that coordination in LLM multi-agent systems may be governed by heterogeneous cascade dynamics. Similar dynamics have been widely observed in other complex systems~\cite{barabasi1999emergence, barabasi2005origin, newman2005power, goh2001universal}.

Motivated by these observations, we propose a testable hypothesis: coordination event sizes may follow heavy-tailed distributions under finite constraints. We empirically validate this hypothesis over \textbf{1.5 million coordination events}, spanning diverse coordination topologies, task families, system scales, and model families. Specifically, we introduce a set of \textit{atomic coordination events}: \texttt{delegation cascades, revision waves, contradiction bursts, synthesis merges, and total cognitive effort}, as observables for capturing how reasoning activity propagates and accumulates. A systematic analysis across all setups indicates that coordination event sizes consistently follow truncated power-law distributions, with estimated exponents $\hat{\alpha} \in (2,3)$. These estimates are obtained via maximum-likelihood estimation following established methodology~\cite{newman2005power, clauset2009power}, and remain within $(2,3)$ across tasks, topologies, scales, and model families. Likelihood-ratio tests under the Vuong framework~\cite{vuong1989likelihood} further confirm that the truncated power law provides a significantly better fit than alternative candidate distributions, such as the log-normal and pure power law, with $p < 0.05$. Importantly, we do not assume power-law behavior a priori. Instead, we statistically test for heavy-tailed structure under finite constraints following established inference procedures~\cite{newman2005power, stumpf2012critical}. To the best of our knowledge, this is the first systematic empirical evidence that coordination in LLM multi-agent systems exhibits heavy-tailed cascade dynamics at the level of reasoning events. These results identify a consistent organizational regime characterized by heterogeneous cascades, with direct implications for how collective reasoning is structured and scaled.

\begin{figure*}[t]
    \centering
    \vspace{-2em}
    \includegraphics[width=\textwidth]{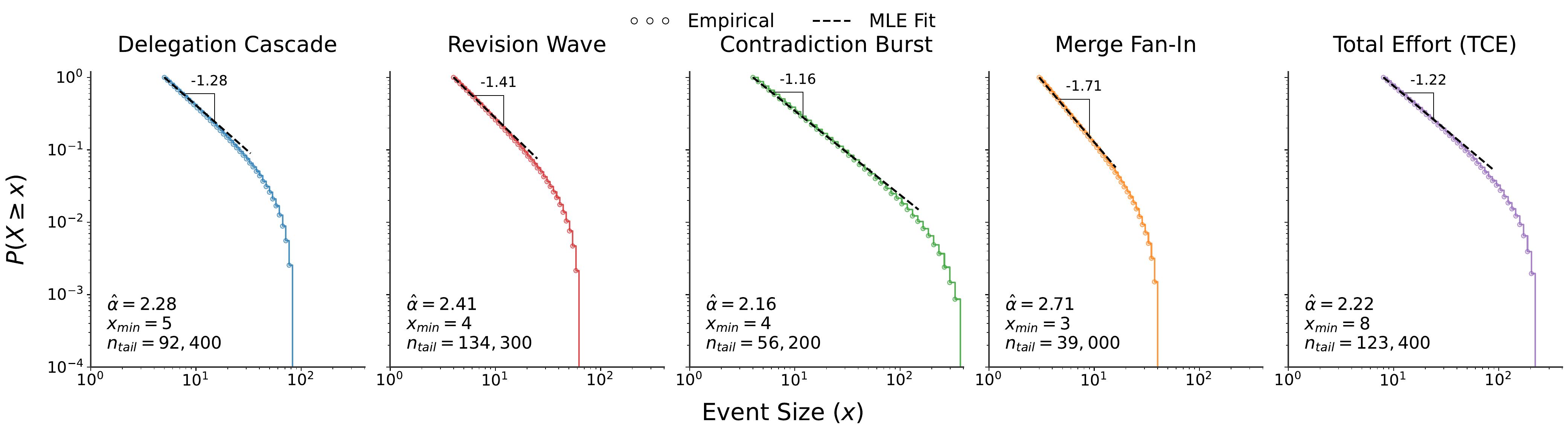}
    \caption{
\textbf{Heavy-tailed coordination cascades across observables.} CCDFs show a power-law regime ($2 < \hat{\alpha} < 3$) with truncation at large $x$. Dashed lines indicate MLE fits above $x_{\min}$. Truncated power laws are favored over log-normal and exponential alternatives (Table~\ref{tab:global_tail_model_comparison}).
}
    \label{fig:ccdf-overall}
    \vspace{-2em}
\end{figure*}

We move beyond the law itself to further investigate its implications. We show that the observed heavy-tailed structure is consistent with reinforcement in coordination: claims that accumulate early engagement tend to attract disproportionately more downstream activity, and this effect strengthens with system size, analogous to preferential attachment in complex systems~\cite{barabasi1999emergence}. This leads to a concentration of cognitive effort in a small subset of agents, forming an emergent tier of \textbf{\texttt{intellectual elites}} whose influence increases with scale. More importantly, the internal composition of large cascades reveals a structural imbalance. As system size increases, cascade generation driven by delegation and contradiction continues to grow, while integration through merge does not scale proportionally. This indicates that although the system becomes increasingly effective at expanding reasoning, it does not become proportionally better at consolidating it. This \textbf{integration bottleneck} provides a mechanistic explanation for the non-monotonic scaling behavior observed in prior work~\cite{chen2024more,cemri2025multi}. Larger coordination cascades are not inherently detrimental. On the contrary, they are essential for exploring complex solution spaces through branching reasoning and task decomposition. The problem is misallocation: as scale increases, a growing fraction of large cascades reflects redundant exploration or unresolved conflict rather than productive synthesis.

This finding suggests that improving LLM MAS requires not only increasing reasoning capacity, but also regulating coordination structure. To examine this, we introduce \textbf{Deficit-Triggered Integration (DTI)}, an intervention that monitors the imbalance between cascade expansion and merge activity within each active cascade and triggers integration when this imbalance exceeds a threshold. DTI still preserves the heavy-tailed regime that supports large-scale reasoning while reducing inefficient late-stage expansion. Across topology-task conditions, DTI consistently improves task success, with the largest gains observed in regimes where the expansion-integration imbalance is most pronounced. These results support the view that coordination failure in LLM MAS is a structural problem and demonstrate that it can be mitigated through targeted regulation of coordination dynamics. More broadly, they point toward a new design paradigm in which scalability depends not only on model capability, but also on how collective reasoning is organized.

\paragraph{We list our contributions below:}
\begin{itemize}
    \item We introduce an event-based formulation of multi-agent reasoning that decomposes coordination into atomic primitives and defines Total Cognitive Effort (TCE) to measure downstream reasoning load.

    \item We examine and validate three consistent observations: \textbf{(H1)} coordination cascades follow truncated power-law distributions, \textbf{(H2)} cognitive effort concentrates in a small subset of agents, and \textbf{(H3)} extreme coordination events scale with system size.

    \item We show that reinforcement amplifies already-engaged reasoning trajectories, leading to intellectual elites, and identify an integration bottleneck where cascade generation outpaces synthesis, explaining non-monotonic scaling behavior.

    \item We propose Deficit-Triggered Integration (DTI), a mechanism derived from observed coordination dynamics that improves performance by regulating the balance between expansion and integration.
\end{itemize}

\section{Related Work}
\label{sec:related}

\paragraph{LLM Multi-Agent Systems and Coordination:} LLM MAS extend reasoning beyond single-agent limits through structured interaction across role-playing, orchestration, software development, and social simulation~\cite{li2023camel,wu2024autogen,hong2023metagpt,qian2024chatdev,park2023generative}. Deliberative mechanisms such as debate improve reasoning quality~\cite{du2024improving,liang2024encouraging,chan2023chateval,chen2024reconcile}, while communication topology strongly shapes collective outcomes~\cite{qian2024scaling,zhuge2024gptswarm,liu2023dynamic}. Despite this progress, scaling agent count does not reliably improve performance and can degrade due to coordination failures and saturation~\cite{chen2024more,kim2025towards,cemri2025multi,kapoor2024ai}. These limitations depend on protocol and scaffold design~\cite{neurips2025scaffolds} and are not fully explained by model capability alone, indicating a lack of principled multi-agent grounding~\cite{la2025large}. Surveys summarize advances and open challenges~\cite{guo2024large,xi2025rise,wang2024survey,tran2025multi}. We evaluate on established agentic benchmarks~\cite{mialon2023gaia,jimenez2023swe,zhu2025multiagentbench,geng2025realm} and position our work relative to single-agent reasoning advances~\cite{wei2022chain,wang2022self,yao2022react,shinn2023reflexion}. However, the statistical structure of coordination dynamics and governing organizational laws remains unexplored, which is the focus of this work.

\paragraph{Collective Intelligence, Inequality, and Elite Formation:} Collective intelligence research shows that groups exhibit emergent cognitive properties beyond individuals and can outperform individual judgment under diversity and independence assumptions~\cite{woolley2010evidence,surowiecki2005wisdom,malone2015handbook}. In multi-agent reinforcement learning, coordinated behavior emerges from local interactions~\cite{leibo2017multi,lowe2017multi}, with recent work extending this view to LLM agent societies~\cite{nisioti2024text}. A parallel literature on inequality provides a lens on intellectual elites. The Lorenz curve, Gini coefficient, and Pareto principle characterize skewed contribution as a structural property of productive systems~\cite{lorenz1905methods,gini1921measurement,pareto1964cours}, while empirical studies show that a small fraction of participants accounts for disproportionate activity, notably in Wikipedia~\cite{halfaker2013rise}. Influence concentration in networks further supports this pattern~\cite{bakshy2011everyone,cha2010measuring,klemm2002highly,goh2001universal}, with broader analyses suggesting such inequality is structurally generated rather than incidental~\cite{piketty2014capital}. However, whether elite formation arises endogenously from coordination cascade dynamics in LLM agent societies, and whether the mechanisms driving heavy-tailed coordination also govern influence concentration, remains unexplored.

\paragraph{Power Laws, Heavy-Tailed Distributions, and Cascade Dynamics:}
Heavy-tailed distributions and scale-free organization are canonical signatures of complex systems, arising from mechanisms such as preferential attachment~\cite{barabasi1999emergence}, small-world structure~\cite{watts1998collective}, and self-organized criticality~\cite{bak1987self}. These dynamics produce power-law event distributions observed across citation networks, information cascades~\cite{newman2003structure,watts2002simple,leskovec2007dynamics}, human activity~\cite{barabasi2005origin}, and social contagion~\cite{crane2008robust}. Distinguishing power-law from log-normal behavior, central to our model comparison, has been extensively studied~\cite{mitzenmacher2004brief,newman2005power}, with rigorous inference requiring careful statistical validation~\cite{stumpf2012critical,goldstein2004problems}. We adopt the Clauset–Shalizi–Newman framework~\cite{clauset2009power} for maximum-likelihood estimation, goodness-of-fit testing, and model comparison, with discrete fitting following Virkar \& Clauset~\cite{virkar2012power} and likelihood-ratio selection via Vuong’s test~\cite{vuong1989likelihood}. Extreme-event scaling is grounded in classical extreme value theory~\cite{embrechts2013modelling}. While these methods are well established, their application to coordination dynamics in LLM MAS and the resulting implications for collective reasoning remains unexplored.

\section{Methodology}
\label{sec:method}

\subsection{Experimental Setup and Data Generation}
\label{subsec:exp-setup}

We study coordination in LLM MAS using structured workloads derived from four SOTA agent benchmarks: GAIA~\cite{mialon2023gaia}, SWE-bench~\cite{jimenez2023swe}, REALM-Bench~\cite{geng2025realm}, and MultiAgentBench~\cite{zhu2025multiagentbench}, spanning QA, reasoning, coding, and planning tasks. Each run consists of $N \in \{8,16,32,64,128,256,512\}$ agents solving interdependent tasks under communication topologies including \textit{chain, star, tree, hierarchical, fully connected, sparse mesh, and dynamic reputation}. We adopt a standardized execution protocol: all agents share a common LLM, prompt, tools, and task instances, with execution implemented in LangGraph~\cite{langgraph2024} to enforce topology and routing (Sec.~\ref{app:agent_config}). To maintain balanced coordination demand as $N$ increases, workloads are scaled using a benchmark-conditioned expansion module that generates task trees with dependency structure but no prescribed coordination, allowing interaction patterns to emerge (Appendix~\ref{app:workload_expansion}). Agents execute over the \textit{task tree} by iteratively selecting, decomposing, and solving tasks while communicating over the active topology. Interaction traces $\mathcal{T}$ are recorded at the level of individual reasoning steps, capturing agent actions, coordination lineage, and task dependencies, forming the basis for analysis (Appendix~\ref{supp-sec:event-formulation}). Experiments are repeated across five seeds per configuration; full details provided in Appendix Section~\ref{app:exp-setup}.

\subsection{Event-Based Coordination Formulation}
\label{subsec:coordination-representation}

To study coordination dynamics in LLM multi-agent systems in a structured manner, we develop an event-based formulation that decomposes coordination into atomic primitives derived from interaction traces. Each trace $\mathcal{T}$ consists of timestamped agent actions with associated references and dependencies, capturing how agents produce outputs and relate them to prior reasoning steps.

\begin{table*}[t]
\centering
\small
\setlength{\tabcolsep}{2.8pt}
\setlength{\extrarowheight}{1.5pt}
\renewcommand{\arraystretch}{1.68}
\arrayrulecolor{gridgray}

\begin{tabular}{|>{\raggedright\arraybackslash}p{2.22cm}
                |>{\centering\arraybackslash}m{3.0cm}
                |>{\raggedright\arraybackslash}p{5.05cm}
                |>{\raggedright\arraybackslash}p{3.00cm}|}
\hline
\rowcolor{headergray}
\textbf{Event Type} & \textbf{Schematic} & \textbf{Definition} & \textbf{What It Captures} \\
\hline

\textbf{Delegation Cascade}
&
\cellcolor{schematicbg}
\rule{0pt}{2.9em}%
\begin{tikzpicture}[scale=0.88, baseline=(current bounding box.center)]
\path[use as bounding box] (-1.45,-1.10) rectangle (1.45,1.55);
\node[evnode, fill=pastelblue] (r) at (0,1.00) {};
\node[evnode, fill=pastelgreen] (a) at (-0.72,0.20) {};
\node[evnode, fill=pastelgreen] (b) at (0.72,0.20) {};
\node[evnode, fill=pastelyellow] (c) at (-1.00,-0.58) {};
\node[evnode, fill=pastelyellow] (d) at (-0.34,-0.58) {};
\draw[evline] (r) -- (a);
\draw[evline] (r) -- (b);
\draw[evline] (a) -- (c);
\draw[evline] (a) -- (d);
\end{tikzpicture}
\rule[-1.1em]{0pt}{0pt}
&
Number of events in the subtask tree rooted at a \texttt{delegate\_subtask} event.
&
Recursive task decomposition and agent recruitment.
\\
\hline

\textbf{Revision Wave}
&
\cellcolor{schematicbg}
\rule{0pt}{2.4em}%
\begin{tikzpicture}[scale=0.88, baseline=(current bounding box.center)]
\path[use as bounding box] (-0.45,-0.65) rectangle (2.95,0.70);
\node[evnode, fill=pastelblue]   (c1) at (0,0) {};
\node[evnode, fill=pastelgreen]  (c2) at (0.82,0) {};
\node[evnode, fill=pastelyellow] (c3) at (1.64,0) {};
\node[evnode, fill=pastelpink]   (c4) at (2.46,0) {};
\draw[evline] (c1) -- (c2);
\draw[evline] (c2) -- (c3);
\draw[evline] (c3) -- (c4);
\end{tikzpicture}
\rule[-0.95em]{0pt}{0pt}
&
Length of a chain of \texttt{revise\_claim} events linked by \texttt{parent\_claim\_id}.
&
Iterative refinement of a claim.
\\
\hline

\textbf{Contradiction Burst}
&
\cellcolor{schematicbg}
\rule{0pt}{2.8em}%
\begin{tikzpicture}[scale=0.88, baseline=(current bounding box.center)]
\path[use as bounding box] (-1.45,-1.05) rectangle (1.45,1.05);
\node[evnode, fill=pastelblue] (p) at (0,0) {};
\node[evnode, fill=pastelpink] (a) at (-1.00,0.68) {};
\node[evnode, fill=pastelpink] (b) at (-1.00,-0.68) {};
\node[evnode, fill=pastelpink] (c) at (1.00,0.68) {};
\node[evnode, fill=pastelpink] (d) at (1.00,-0.68) {};
\draw[evline] (a) -- (p);
\draw[evline] (b) -- (p);
\draw[evline] (c) -- (p);
\draw[evline] (d) -- (p);
\end{tikzpicture}
\rule[-1.0em]{0pt}{0pt}
&
Number of distinct agents issuing \texttt{contradict\_claim} on the same parent claim.
&
Parallel critique centered on one claim.
\\
\hline

\textbf{Merge Fan-in}
&
\cellcolor{schematicbg}
\rule{0pt}{2.7em}%
\begin{tikzpicture}[scale=0.88, baseline=(current bounding box.center)]
\path[use as bounding box] (-1.55,-1.00) rectangle (1.55,1.00);
\node[evnode, fill=pastelgreen]  (a) at (-1.10,0.62) {};
\node[evnode, fill=pastelyellow] (b) at (-1.10,-0.62) {};
\node[evnode, fill=pastelpink]   (c) at (-0.30,0.10) {};
\node[evnode, fill=pastelpurple] (m) at (1.10,0) {};
\draw[evline] (a) -- (m);
\draw[evline] (b) -- (m);
\draw[evline] (c) -- (m);
\end{tikzpicture}
\rule[-1.0em]{0pt}{0pt}
&
Number of \texttt{parent\_claim\_ids} referenced by a single \texttt{merge\_claims} event.
&
Information integration bottleneck.
\\
\hline

\textbf{Total Cognitive Effort (TCE)}
&
\cellcolor{schematicbg}
\rule{0pt}{2.9em}%
\begin{tikzpicture}[scale=0.88, baseline=(current bounding box.center)]
\path[use as bounding box] (-1.45,-1.05) rectangle (1.45,1.50);
\node[evnode, fill=pastelblue]   (r) at (0,1.00) {};
\node[evnode, fill=pastelgreen]  (a) at (-1.00,0.18) {};
\node[evnode, fill=pastelyellow] (b) at (0,0.18) {};
\node[evnode, fill=pastelpink]   (c) at (1.00,0.18) {};
\node[evnode, fill=pastelyellow] (d) at (-0.46,-0.66) {};
\node[evnode, fill=pastelgreen]  (e) at (0.46,-0.66) {};
\draw[evline] (r) -- (a);
\draw[evline] (r) -- (b);
\draw[evline] (r) -- (c);
\draw[evline] (b) -- (d);
\draw[evline] (c) -- (e);
\end{tikzpicture}
\rule[-1.1em]{0pt}{0pt}
&
Total number of downstream coordination events linked to a root claim.
&
Aggregate cascade size.
\\
\hline
\end{tabular}

\vspace{6pt} 
\caption{\textbf{Primitive coordination events used in our analysis.}
An \emph{event} is a single coordination step in the reasoning process (e.g., delegation, revision, contradiction, or merge). 
Events are linked through claim and subtask relationships, and each row defines a quantity computed over these events rather than over entire tasks.
\textbf{Legend:} 
\legenditem{pastelblue}{Root Claim} \quad
\textcolor{pastelgreen}{\rule{0.9em}{0.9em}}\!\textcolor{pastelyellow}{\rule{0.9em}{0.9em}}\hspace{0.3em}\textit{Propagation Claims} \quad
\legenditem{pastelpink}{Critique Claim} \quad
\legenditem{pastelpurple}{Merge Claim}.}
\label{tab:event-definition-table} 

\end{table*}


We define a \textit{claim} as the atomic unit of reasoning produced by an agent at a given step, corresponding to an intermediate output during task execution. Let $\mathcal{C}$ denote the set of all claims. Claims are uniquely identified and linked to prior claims through recorded references, inducing a directed acyclic graph (DAG) $\mathcal{G} = (\mathcal{C}, \mathcal{E}_c)$, where $(c_i, c_j) \in \mathcal{E}_c$ if $c_i \in \mathcal{P}(c_j)$. $\mathcal{P}(c_j) \subseteq \mathcal{C}$ denotes the set of parent claims referenced by $c_j$. This graph captures the evolution of reasoning across agents, tasks, and interactions.

An \textit{event} is defined as a coordination step corresponding to a recorded action that transforms or relates one or more claims, identified via the action type and reference structure. Let $\mathcal{E}$ denote the set of all events. These events represent atomic coordination primitives, including decomposition, refinement, critique, synthesis, and reuse.

Each claim is associated with a root identifier, which defines a \textit{cascade} as the set of all claims that share the same root. Formally, for a root claim $c_r$, the corresponding cascade is given by
\[
\mathcal{C}_r = \{c_i \in \mathcal{C} \mid \text{root}(c_i) = c_r\}.
\]
Let $\mathcal{E}_r \subseteq \mathcal{E}$ denote the set of events associated with cascade $\mathcal{C}_r$. Cascades therefore represent connected subgraphs of $\mathcal{G}$ corresponding to the propagation of reasoning initiated by a single claim, forming the fundamental units for analyzing coordination dynamics.

Detailed notation, definitions of claim types, event categories, and graph construction are provided in Appendix~\ref{supp-sec:event-formulation} and~\ref{supp-sec:notations}.

\subsection{Observables}

We define a set of observables over the claim graph and cascades to quantify coordination dynamics. Let $\mathcal{G} = (\mathcal{C}, \mathcal{E}_c)$ denote the claim graph and $\{\mathcal{C}_r\}$ the set of cascades as defined in Sec.~\ref{subsec:coordination-representation}. Observables are computed at the level of individual events, claims, and cascades, enabling analysis of coordination behavior across multiple scales.

\paragraph{Event size:} We first define the size of a coordination event as the number of claims involved in the corresponding transformation. For a given event $e_k$, we denote its size by $x(e_k)$, which depends on the event type. We define \textit{four} event types: \texttt{delegation, revision, merge fan-in and contradiction}, whose definitions are provided in Table~\ref{tab:event-definition-table}. These event sizes form the primary units for distributional analysis.

\paragraph{Cascade size:} The size of a cascade at a root claim $c_r$ is defined as the total number of claims associated with a root claim. This is given by:

\begin{equation}
    |\mathcal{C}_r| = \sum_{c_i \in \mathcal{C}_r} 1
    \label{eq:cascade-size}
\end{equation}

\paragraph{Total Cognitive Effort (TCE):} Building on this, we define the TCE of a cascade as the total number of coordination events generated within that cascade:

\begin{equation}
\text{TCE}(c_r) = \sum_{e_k \in \mathcal{E}_r} 1
\label{eq:tce}
\end{equation}

where $\mathcal{E}_r$ denotes the set of events associated with claims in $\mathcal{C}_r$. TCE captures the total amount of reasoning activity triggered by a single root claim, and serves as a central observable for characterizing coordination complexity.

\paragraph{Contribution concentration:} To analyze contribution concentration, we define agent-level participation within a cascade. Let $n_a(c_r)$ denote the number of claims produced by agent $a$ within cascade $\mathcal{C}_r$. We define the cumulative contribution share of the top-$k$ agents as

\begin{equation}
    S_k(c_r) = \frac{\sum_{a \in \text{Top-}k} n_a(c_r)}{\sum_{a \in \mathcal{A}} n_a(c_r)}
\end{equation}

This measure quantifies the degree to which coordination effort is concentrated among a subset of agents. In subsequent analysis, we instantiate this as $E^{\text{active}}_k$ when  the denominator is restricted to active agents only, and $E^{\text{all}}_k$ when computed over all $N$ agents. These are the primary concentration  observables reported in figures and tables throughout the paper.


\paragraph{Contribution concentration:} Let $n_a(c_r)$ denote the number of claims produced by agent $a$ in cascade $\mathcal{C}_r$. The top-$k$ contribution share is:
$S_k(c_r) = \frac{\sum_{a \in \text{Top-}k} n_a(c_r)}{\sum_{a \in \mathcal{A}} n_a(c_r)}$.

\paragraph{Extreme-event scaling:} For a given agent population $N$, we define the maximum cascade size as, $x_{\max}(N) = \max_{c_r} |\mathcal{C}_r|$.

These observables characterize the distributional structure, concentration, and scaling behavior of coordination in LLM multi-agent systems. In subsequent analysis, we report $S_k$ as $E^{\mathrm{active}}_{k}$ when computed over active agents only, and $E^{\mathrm{all}}_{k}$ when computed over all $N$ agents. Please see Appendix Sec.~\ref{supp-sec:notations} for full lists of notation.

\subsection{Statistical Methods}

We analyze the distributional properties of coordination observables using established statistical methods for heavy-tailed data. For each observable (e.g., event size, cascade size, and TCE), we estimate candidate distributions including the power law, truncated power law, log-normal, and exponential. Parameter estimation is performed using maximum likelihood estimation (MLE). For heavy-tailed models, the lower cutoff $x_{\min}$ is selected by minimizing the Kolmogorov–Smirnov (KS)~\cite{an1933sulla, smirnov1948table} distance between the empirical distribution and the fitted model. All distributions are fit over the tail region $x \geq x_{\min}$. To compare competing models, we use likelihood ratio tests following Vuong’s method for non-nested distributions~\cite{vuong1989likelihood}. For each pair of candidate models, we report the log-likelihood ratio and its statistical significance. Goodness-of-fit is evaluated using the KS statistic, and distributional behavior is visualized using complementary cumulative distribution functions (CCDFs) on log–log axes~\cite{newman2005power}. This approach avoids biases associated with binning and provides a stable representation of tail behavior. All statistical analyses are implemented using the \texttt{powerlaw} package, following the methodology of Clauset et al.~\cite{clauset2009power} and subsequent best practices for heavy-tailed inference. Additional statistical metrics and reliability tests are provided in Sec.~\ref{app:quantitative}.

\section{Theory of Reinforced Routing}
\label{sec:theory}

\subsection{Overview: Recursive Coordination and Event Propagation}

Coordination in LLM multi-agent systems unfolds as a recursive process over claims. At each step, agents operate on existing claims to produce new ones through actions such as decomposition, refinement, critique, and synthesis. This creates a dependency structure in which reasoning propagates through a growing network of interrelated claims rather than being generated independently. Once a root claim is introduced, it can trigger a cascade of downstream activity. Delegation expands the reasoning structure, revision refines intermediate outputs, contradiction creates branching alternatives, and merge operations integrate multiple reasoning paths. These interactions give rise to cascades in which claims recursively generate further claims. This process is inherently path-dependent: claims that accumulate more interaction become more likely to be revisited and expanded, while others remain inactive. As a result, coordination concentrates around a subset of evolving reasoning trajectories. Understanding how this concentration emerges motivates a mechanism for how agents select and build upon existing claims.

\subsection{Reinforced Routing Mechanism}

We model coordination in LLM multi-agent systems as a routing process over claims. Let $c_i$ denote a claim active at time $t$, and let $x_i(t)$ denote its accumulated coordination activity, defined as the number of downstream coordination events that reference, revise, contradict, merge from, or delegate from $c_i$ up to time $t$. This quantity captures the extent to which a claim has been involved in the ongoing reasoning process.

Agents repeatedly select existing claims as the basis for further reasoning. This selection is not uniform: claims that have accumulated more activity are more likely to be revisited and expanded, reflecting their prominence in the shared reasoning state and their relevance to task progress. We formalize this behavior through a reinforced routing rule:
\begin{equation}
\mathbb{P}(c_i \mid \mathcal{F}_t)
=
\frac{x_i(t)^{\beta}}{\sum_j x_j(t)^{\beta}},
\label{eq:reinforced_routing}
\end{equation}
where $\mathcal{F}_t$ denotes the interaction history and $\beta > 0$ controls the strength of reinforcement. When $\beta = 0$, routing is uniform over claims; when $\beta > 0$, already-active claims are preferentially selected, with larger $\beta$ corresponding to stronger amplification of activity.

A direct implication of this mechanism is that the continuation probability of a claim increases with its prior engagement. Let $R(x,N)$ denote the routing ratio for claims with activity $x$ in a system of size $N$, measured relative to an activity-independent baseline. In the reinforcement regime,
\begin{equation}
R(x,N) \propto x^{\beta(N)}
\label{eq:pref-attachment}
\end{equation}
where the preferential attachment exponent is defined as
\begin{equation}
\beta(N) = \frac{d \log R(x,N)}{d \log x}.
\label{eq:beta_definition}
\end{equation}.

In isolation, reinforced routing generates broad, scale-free coordination patterns. However, LLM multi-agent systems are intrinsically bounded by finite agent populations, context limits, communication overhead, and constrained task depth. These constraints suppress unbounded cascade growth and truncate the tail of the induced distribution. We therefore model coordination observables using a truncated power law:
\begin{equation}
P(X = x) \propto x^{-\alpha} e^{-x/x_c}
\label{eq:tpl}
\end{equation}
where $\alpha$ governs the intermediate scaling regime and $x_c$ denotes the characteristic cutoff imposed by system constraints.

\subsection{Cascade Growth Model}

Under reinforced routing, coordination propagates through claim-rooted cascades. Each selected claim can generate new claims via delegation, revision, contradiction, or synthesis, inducing a branching process over the claim graph. Let $\lambda$ denote the expected number of new claims generated per selected claim. When $\lambda < 1$, cascades remain small; when $\lambda > 1$, they grow rapidly; and when $\lambda \approx 1$, cascade sizes exhibit high variability across scales. Reinforcement amplifies this process by increasing the likelihood that already-active claims continue to generate further activity. Cascades that gain early momentum are therefore more likely to persist and expand, while others terminate quickly, producing heterogeneous growth across cascades. However, cascade expansion is bounded by finite system constraints, including limited agents, context, communication capacity, and task depth. These constraints cap the effective branching process and prevent unbounded growth. As a result, coordination operates near criticality but remains finite, yielding broad but truncated cascade size distributions.

\subsection{Theoretical Implications}

Reinforced routing combined with bounded cascade growth yields characteristic coordination behavior in LLM multi-agent systems. Agents iteratively build on intermediate reasoning outputs within a shared interaction state, concentrating activity on a subset of claims while coordination expands through cascades under finite computational and communication constraints. \textbf{\ding{192} Heavy-tailed coordination.} Reinforcement amplifies activity along selected reasoning trajectories, producing large variation in event sizes and cascade magnitudes. Under bounded resources such as limited context, token budgets, and coordination overhead, this yields heavy-tailed but truncated distributions, as described by Eq.~\eqref{eq:tpl}. \textbf{\ding{193} Emergence of intellectual elites.} Because agents preferentially operate on already-active claims, a subset of reasoning trajectories attracts disproportionate coordination. Agents contributing to these trajectories participate in larger cascades and accumulate greater influence, leading to the emergence of intellectual elites despite symmetric initial capabilities. \textbf{\ding{194} Scaling of extremes.} As the number of agents increases, the volume of concurrent reasoning interactions grows, increasing the likelihood of large cascades. This results in systematic growth of the largest coordination events with $N$. \textbf{\ding{195} Expansion–integration imbalance.} Coordination primitives such as delegation, revision, and critique drive rapid expansion of reasoning branches, while integration through synthesis remains comparatively constrained. As cascades grow, this imbalance produces increasingly complex but weakly integrated reasoning structures. Together, these implications show that heavy-tailed coordination, elite formation, and scaling behavior arise naturally from reinforced and bounded coordination dynamics in LLM multi-agent systems.

\section{Empirical Laws of Collective Cognition}
\label{sec:empirical-laws}

We study coordination in LLM multi-agent systems through three hypotheses on the statistical structure, organization, and scaling of collective reasoning, evaluated across scales, topologies, task domains, and coordination primitives to uncover underlying mechanisms.

\textbf{\ding{182} H1.} Coordination in LLM multi-agent systems exhibits heavy-tailed cascade structure under finite system constraints.

\textbf{\ding{183} H2.} Collective reasoning self-organizes into unequal contribution regimes, leading to the emergence of intellectual elites.

\textbf{\ding{184} H3.} Extreme coordination cascades grow systematically with agent society size.

\subsection{H1: Heavy-Tailed Coordination Cascades}

We test whether coordination in LLM MAS organizes through bounded heavy-tailed cascades by analyzing total cognitive effort (TCE) and its constituent coordination primitives across event types, task domains, topologies, and scales.

\begin{table*}[h]
\centering
\scriptsize
\setlength{\tabcolsep}{4pt}
\renewcommand{\arraystretch}{1.05}

\resizebox{\textwidth}{!}{
\begin{tabular}{lrrrrlrr|rrrr}
\toprule
& \multicolumn{7}{c|}{\textbf{Tail statistics}} & \multicolumn{4}{c}{\textbf{Model comparison}} \\
\cmidrule(r){2-8} \cmidrule(l){9-12}

\textbf{Observable}
& $n_{\text{total}}$
& $x_{\min}$
& $n_{\text{tail}}$
& $\hat{\alpha}$
& Tail family
& $x_{\max}$
& Distinct
& LR (TPL vs LN)
& $p$
& LR (TPL vs PL)
& $p$ \\

\midrule

Delegation cascade
& 342{,}300 & 5 & 92{,}400 & 2.28 & Trunc.\ PL & 118 & 103
& +4.6 & 0.002 & +2.8 & 0.009 \\

Revision wave
& 344{,}300 & 4 & 134{,}300 & 2.41 & Trunc.\ PL & 74 & 61
& +3.8 & 0.005 & +1.7 & 0.037 \\

Contradiction burst
& 255{,}300 & 4 & 56{,}200 & 2.16 & Trunc.\ PL & 83 & 73
& +4.2 & 0.004 & +0.9 & 0.020 \\

Merge fan-in
& 278{,}500 & 3 & 39{,}000 & 2.71 & Trunc.\ PL & 33 & 27
& +3.9 & 0.006 & +3.3 & 0.001 \\

Total Cognitive Effort (TCE)
& 411{,}300 & 8 & 123{,}400 & 2.22 & Trunc.\ PL & 1320 & 1203
& +5.1 & 0.001 & +2.8 & 0.012 \\

\bottomrule
\end{tabular}
}
\caption{
\textbf{Global tail statistics and model comparisons.}
Tail behavior is summarized by $\hat{\alpha}$ (MLE above $x_{\min}$) and distinct tail values. Positive likelihood ratios favor TPL over LN and PL. \textit{* Lower $p$ indicates stronger support for TPL.}
}
\label{tab:global_tail_model_comparison}
\end{table*}

\begin{figure*}[t]
    \centering
    \includegraphics[width=\textwidth]{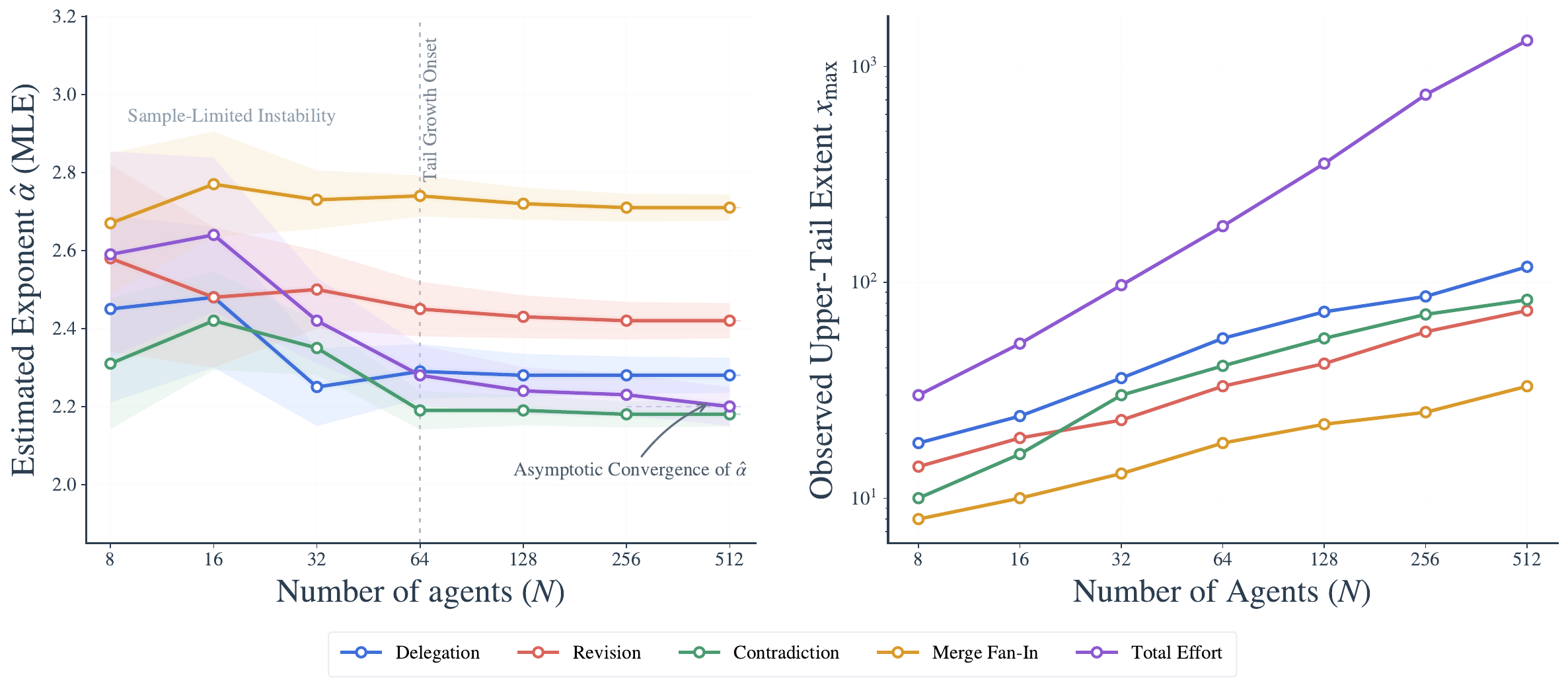}
    \caption{
\textbf{Finite-size stability of heavy-tailed coordination dynamics.}
\textbf{(Left)} Estimated tail exponents $\hat{\alpha}$ (MLE) vs.\ agent count $N$. Estimates fluctuate at small $N$ due to limited tail samples, then stabilize and converge beyond $N \approx 64$, indicating emergence of a consistent heavy-tailed regime.  \textbf{(Right)} Mean maximum event size $\langle x_{\max} \rangle$ vs.\ $N$. The upper tail grows systematically across observables, with strongest expansion for TCE, showing that increasing system size expands the reachable coordination tail.
}
    \label{fig:scaling-with-N}
    \vspace{-7pt}
\end{figure*}

Figure~\ref{fig:ccdf-overall} and Table~\ref{tab:global_tail_model_comparison} establish the primary evidence. CCDFs across all observables exhibit an extended log-log linear region followed by systematic truncation in the far tail. Likelihood-ratio tests favor truncated power-law (TPL) models over both power-law (PL) and log-normal (LN) alternatives ($p < 0.05$), with $\hat{\alpha} \in (2,3)$ and finite cutoff $\hat{x}_c$. This supports Eq.~\ref{eq:tpl}, indicating power-law amplification over an intermediate range with a structural cutoff imposed by context limits, token budgets, and coordination overhead. Coordination is therefore inherently uneven: most trajectories remain local, while a small fraction accumulates disproportionately large activity. Figure~\ref{fig:scaling-with-N} (left) shows that this behavior persists with scale, with $\hat{\alpha}$ stabilizing by $N \approx 64$ and remaining consistent as $N$ increases.

The structure of the tail reveals how this heterogeneity is generated. Delegation and contradiction extend deepest (Fig.~\ref{fig:ccdf-overall}), while merge fan-in is sharply truncated. This follows from their functional roles: delegation and contradiction expand the reasoning space by creating new branches and recursive dependencies, whereas merge is integrative and does not generate new coordination structures. Cascade growth, therefore, proceeds through repeated expansion and branching, rather than through aggregation.

\begin{table*}[h]
\centering
\scriptsize
\setlength{\tabcolsep}{3pt}
\renewcommand{\arraystretch}{1.06}
\resizebox{\textwidth}{!}{
\begin{tabular}{ll|cccc|cccc|cccc|cccc}
\toprule
& & \multicolumn{4}{c|}{\textbf{Chain}}
  & \multicolumn{4}{c|}{\textbf{Star}}
  & \multicolumn{4}{c|}{\textbf{Hierarchical}}
  & \multicolumn{4}{c}{\textbf{Mesh}} \\
\cmidrule(r){3-6}\cmidrule(r){7-10}\cmidrule(r){11-14}\cmidrule(l){15-18}
\textbf{Observable} & \textbf{Task}
& $\hat{\alpha}$ & $\hat{x}_c$ & LR$_{T/LN}$ & LR$_{T/PL}$
& $\hat{\alpha}$ & $\hat{x}_c$ & LR$_{T/LN}$ & LR$_{T/PL}$
& $\hat{\alpha}$ & $\hat{x}_c$ & LR$_{T/LN}$ & LR$_{T/PL}$
& $\hat{\alpha}$ & $\hat{x}_c$ & LR$_{T/LN}$ & LR$_{T/PL}$ \\
\midrule
\multirow{4}{*}{Delegation}
& Coding
& 2.46 & 22 & $+4.0$ & $+2.5$
& 2.30 & 29 & $+4.3$ & $+2.7$
& 2.36 & 26 & $+4.2$ & $+2.6$
& 2.16 & 37 & $+4.6$ & $+2.8$ \\
& QA
& 2.64 & 15 & $+3.7$ & $+2.3$
& 2.48 & 19 & $+3.9$ & $+2.5$
& 2.54 & 17 & $+3.8$ & $+2.4$
& 2.34 & 24 & $+4.2$ & $+2.6$ \\
& Planning
& 2.24 & 36 & $+4.5$ & $+2.8$
& 2.08 & 48 & $+4.9$ & $+3.0$
& 2.14 & 43 & $+4.8$ & $+2.9$
& 1.94 & 61 & $+5.3$ & $+3.2$ \\
& Reasoning
& 2.34 & 29 & $+4.3$ & $+2.7$
& 2.18 & 39 & $+4.7$ & $+2.8$
& 2.24 & 35 & $+4.5$ & $+2.8$
& 2.04 & 49 & $+5.0$ & $+3.0$ \\
\midrule
\multirow{4}{*}{Merge Fan-In}
& Coding
& 2.89 &  6 & $+3.3$ & $+3.0$
& 2.73 &  8 & $+3.6$ & $+3.2$
& 2.79 &  7 & $+3.5$ & $+3.1$
& 2.59 & 10 & $+3.9$ & $+3.3$ \\
& QA
& 3.07 &  4 & $+2.9$ & $+2.8$
& 2.91 &  5 & $+3.2$ & $+3.0$
& 2.97 &  5 & $+3.2$ & $+2.9$
& 2.77 &  7 & $+3.5$ & $+3.1$ \\
& Planning
& 2.67 & 10 & $+3.8$ & $+3.3$
& 2.51 & 13 & $+4.2$ & $+3.5$
& 2.57 & 12 & $+4.0$ & $+3.4$
& 2.37 & 17 & $+4.6$ & $+3.7$ \\
& Reasoning
& 2.77 &  8 & $+3.6$ & $+3.2$
& 2.61 & 11 & $+4.0$ & $+3.3$
& 2.67 & 10 & $+3.8$ & $+3.3$
& 2.47 & 14 & $+4.3$ & $+3.5$ \\
\midrule
\multirow{4}{*}{Revision}
& Coding
& 2.59 & 13 & $+3.2$ & $+1.4$
& 2.43 & 18 & $+3.5$ & $+1.6$
& 2.49 & 16 & $+3.4$ & $+1.5$
& 2.29 & 23 & $+3.8$ & $+1.7$ \\
& QA
& 2.77 &  9 & $+2.9$ & $+1.2$
& 2.61 & 12 & $+3.2$ & $+1.4$
& 2.67 & 11 & $+3.0$ & $+1.3$
& 2.47 & 15 & $+3.4$ & $+1.5$ \\
& Planning
& 2.37 & 22 & $+3.7$ & $+1.7$
& 2.21 & 30 & $+4.1$ & $+1.9$
& 2.27 & 27 & $+4.0$ & $+1.8$
& 2.07 & 38 & $+4.5$ & $+2.1$ \\
& Reasoning
& 2.47 & 18 & $+3.5$ & $+1.6$
& 2.31 & 25 & $+3.9$ & $+1.7$
& 2.37 & 22 & $+3.7$ & $+1.7$
& 2.17 & 31 & $+4.2$ & $+1.9$ \\
\midrule
\multirow{4}{*}{Contradiction}
& Coding
& 2.34 & 16 & $+3.6$ & $+0.6$
& 2.18 & 21 & $+3.9$ & $+0.8$
& 2.24 & 18 & $+3.8$ & $+0.7$
& 2.04 & 26 & $+4.2$ & $+0.9$ \\
& QA
& 2.52 & 11 & $+3.3$ & $+0.4$
& 2.36 & 14 & $+3.6$ & $+0.6$
& 2.42 & 12 & $+3.4$ & $+0.5$
& 2.22 & 17 & $+3.8$ & $+0.7$ \\
& Planning
& 2.12 & 26 & $+4.1$ & $+0.9$
& 1.96 & 34 & $+4.5$ & $+1.1$
& 2.02 & 30 & $+4.4$ & $+1.0$
& 1.82 & 43 & $+4.9$ & $+1.3$ \\
& Reasoning
& 2.22 & 21 & $+3.9$ & $+0.8$
& 2.06 & 28 & $+4.3$ & $+0.9$
& 2.12 & 24 & $+4.1$ & $+0.9$
& 1.92 & 34 & $+4.6$ & $+1.1$ \\
\bottomrule
\end{tabular}
}
\caption{%
\textbf{Tail heterogeneity across task types and communication topologies.}
For each observable and task--topology condition, we report the fitted
truncated power-law exponent $\hat{\alpha}$, cutoff parameter $\hat{x}_c$,
and log-likelihood ratios LR$_{T/LN}$ and LR$_{T/PL}$ comparing the
truncated power law (TPL) against the log-normal (LN) and pure power-law
(PL) alternatives, respectively.
Lower $\hat{\alpha}$ and larger $\hat{x}_c$ indicate broader, less sharply
truncated coordination tails.
Planning and reasoning tasks consistently exhibit heavier tails than coding
and QA across all topology conditions, and denser topologies (mesh, star)
support longer-lived cascades than locally connected structures (chain),
with the planning$\times$mesh combination producing the heaviest tails
in every observable.
}
\label{tab:heterogeneity_drv}
\vspace{-12pt}
\end{table*}

\begin{figure*}[t]
    \centering
    \includegraphics[width=\textwidth]{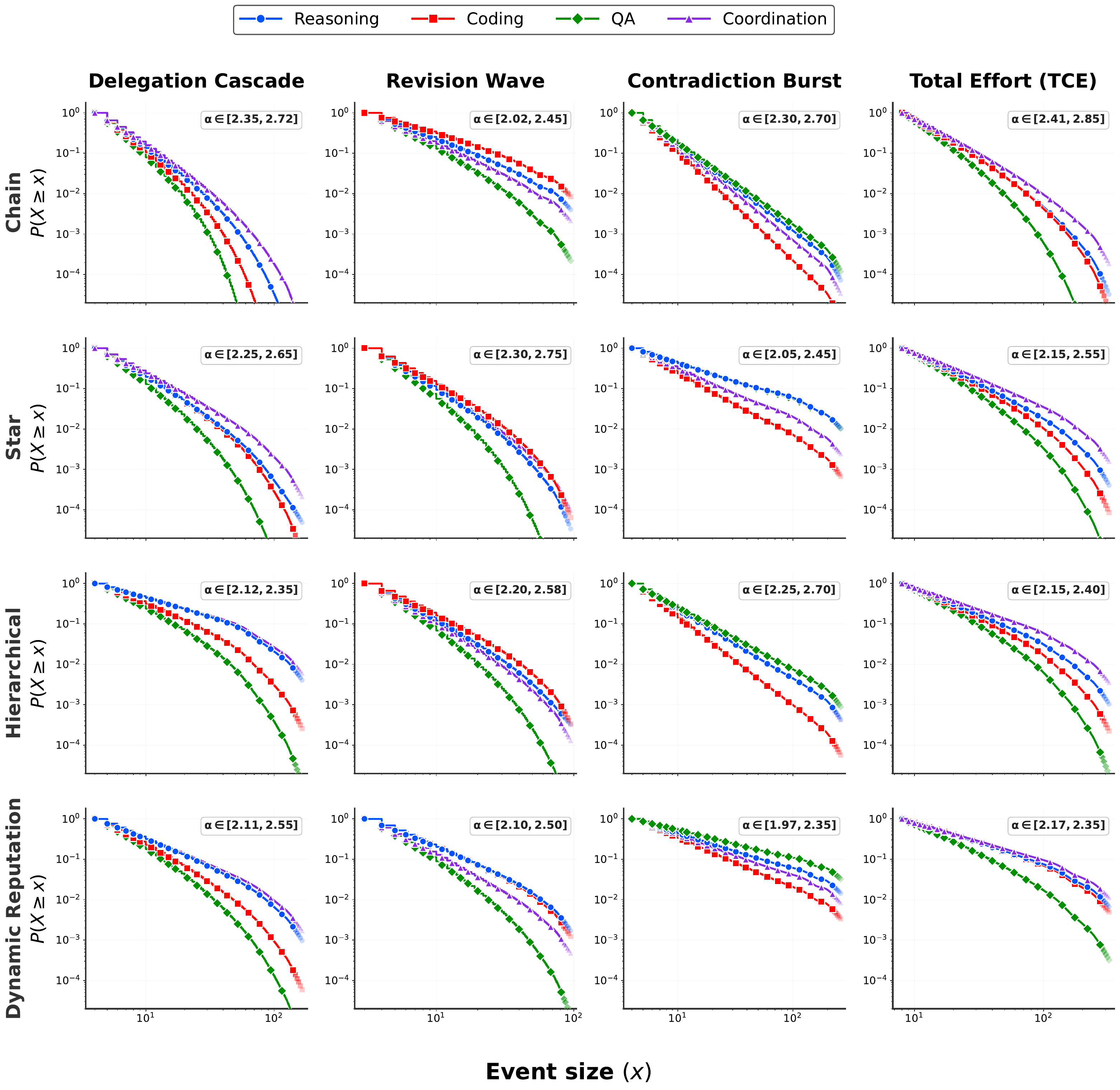}
    \caption{
    \textbf{Topology- and task-specific heavy-tailed coordination cascades in multi-agent LLM systems.}
    Complementary cumulative distribution functions (CCDF) of coordination-event sizes $P(X \ge x)$ across four coordination observables: \textit{Delegation Cascade}, \textit{Revision Wave}, \textit{Contradiction Burst}, and \textit{Total Coordination Effort (TCE)} under four agent interaction topologies (\textit{Chain}, \textit{Star}, \textit{Hierarchical}, and \textit{Dynamic Reputation}) and four task families (\textit{Reasoning}, \textit{Coding}, \textit{QA}, and \textit{Coordination}). Each curve represents the empirical distribution of coordination-event sizes for a given task family within a topology. Across all settings, the distributions exhibit broad heavy-tailed behavior with estimated scaling exponents $2 < \hat{\alpha} < 3$ in the intermediate regime (values shown per panel), consistent with scale-free coordination dynamics. While the precise tail exponent varies modestly across tasks and architectures, the heavy-tailed form persists across all topologies, indicating that complex coordination in LLM multi-agent systems produces heterogeneous cascades spanning multiple scales. Deviations at the largest event sizes reflect finite-size truncation due to system constraints such as limited agent attention, bounded communication bandwidth, and task decomposition depth.
}
    \label{fig:ccdf-task-topology}
    \vspace{-13pt}
\end{figure*}

This mechanism interacts jointly with task structure and topology. As shown in Fig.~\ref{fig:ccdf-task-topology}, tasks with deeper claim dependency graphs (e.g., planning) provide more opportunities for recursive expansion, producing heavier delegation and contradiction tails, while merge remains comparatively insensitive to task depth. Topology determines how far these expansions propagate: denser interaction graphs enable repeated engagement with active trajectories, extending cascades further, whereas sparse structures restrict propagation. The largest cascades arise when these factors align; deep task structure combined with high interaction bandwidth, while as $N$ increases the observable cascade range expands without altering the bounded heavy-tailed form or the TPL preference.

Across model families (Table~\ref{tab:llm-ablation-global}), stronger models exhibit lower $\hat{\alpha}$ and larger $\hat{x}_c$, operating further into the tail, but the TPL form itself is invariant. The task-topology based heterogeneities across different model families show persistent patterns too (Tables~\ref{tab:llm-ablation-tasktype} and \ref{tab:llm-ablation-topology}).

H1 is supported: coordination in LLM multi-agent systems follows a bounded heavy-tailed cascade structure, in which cascade size is governed by the joint interaction of task decomposability, interaction topology, system scale, and coordination primitives.

\subsection{H2: Emergence of Intellectual Elites}

We test whether coordination self-organizes into unequal contribution patterns by analyzing how coordination effort distributes across agents and how this distribution evolves with scale, task structure, and topology.

\begin{figure*}[h]
    \centering
    \includegraphics[width=\textwidth]{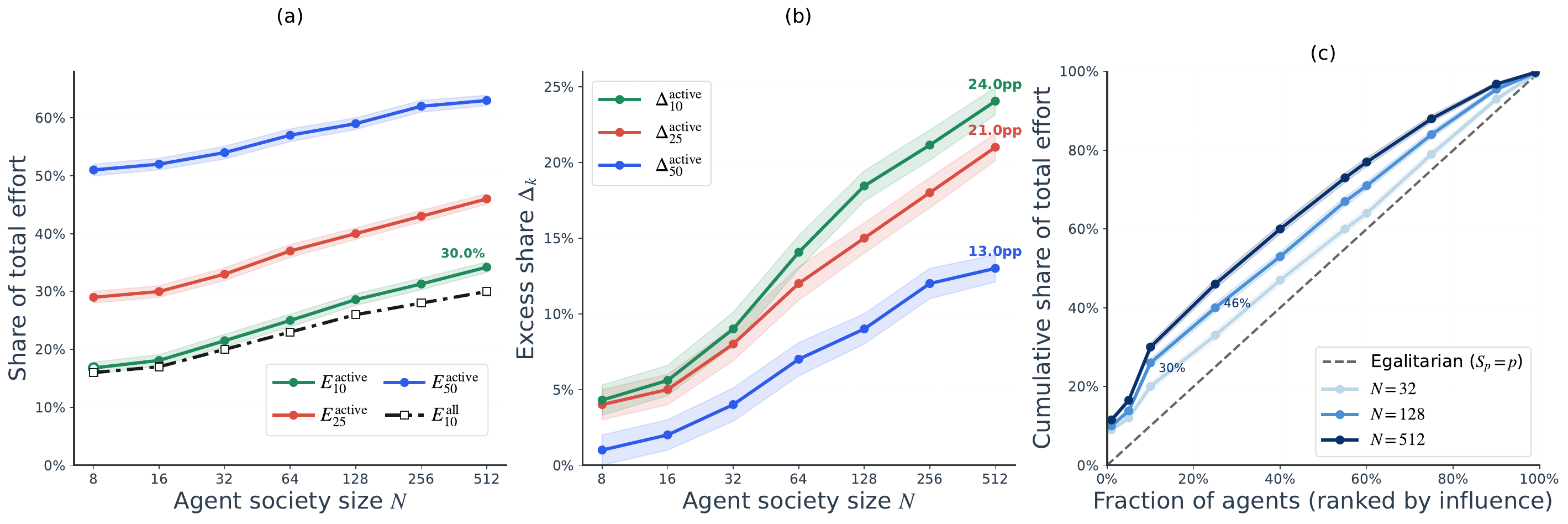}
    \caption{
\textbf{Scale-dependent emergence of broad elite tiers.}
\textbf{(a)} \textbf{(a)} Top-$k$ active agents ($E^{\mathrm{active}}_{10}$, $E^{\mathrm{active}}_{25}$, $E^{\mathrm{active}}_{50}$) capture disproportionate shares of coordination effort relative to egalitarian baselines; $E^{\mathrm{all}}_{10}$ (dashed) confirms the result is not driven by inactive agents. of coordination effort relative to egalitarian baselines.
\textbf{(b)} Excess concentration $\Delta^{\mathrm{active}}_k$ above equal participation increases with $N$, with strongest gains in the top decile and quartile.
\textbf{(c)} Cumulative concentration curves vs $N$ increasingly bow above the equality line, indicating broader and more dominant elite tiers at scale.
}
    \label{fig:elite-global-panel}
    \vspace{-13pt}
\end{figure*}

Figure~\ref{fig:elite-global-panel} provides the primary evidence for concentration. The effort shares of the top-$k$ active agents, $E^{\mathrm{active}}_{10}$, $E^{\mathrm{active}}_{25}$, and $E^{\mathrm{active}}_{50}$, lie well above their egalitarian baselines across all scales, and this gap widens systematically with $N$: the top-10\% excess reaches +24pp at large $N$. The cumulative concentration curves $S_p$ become increasingly convex, showing that larger societies develop broader and more dominant elite tiers rather than converging toward uniform participation. Elite formation is therefore not a finite-size artifact but a scale-amplified structural property.

\begin{figure*}[h]
    \centering
    \includegraphics[width=\textwidth]{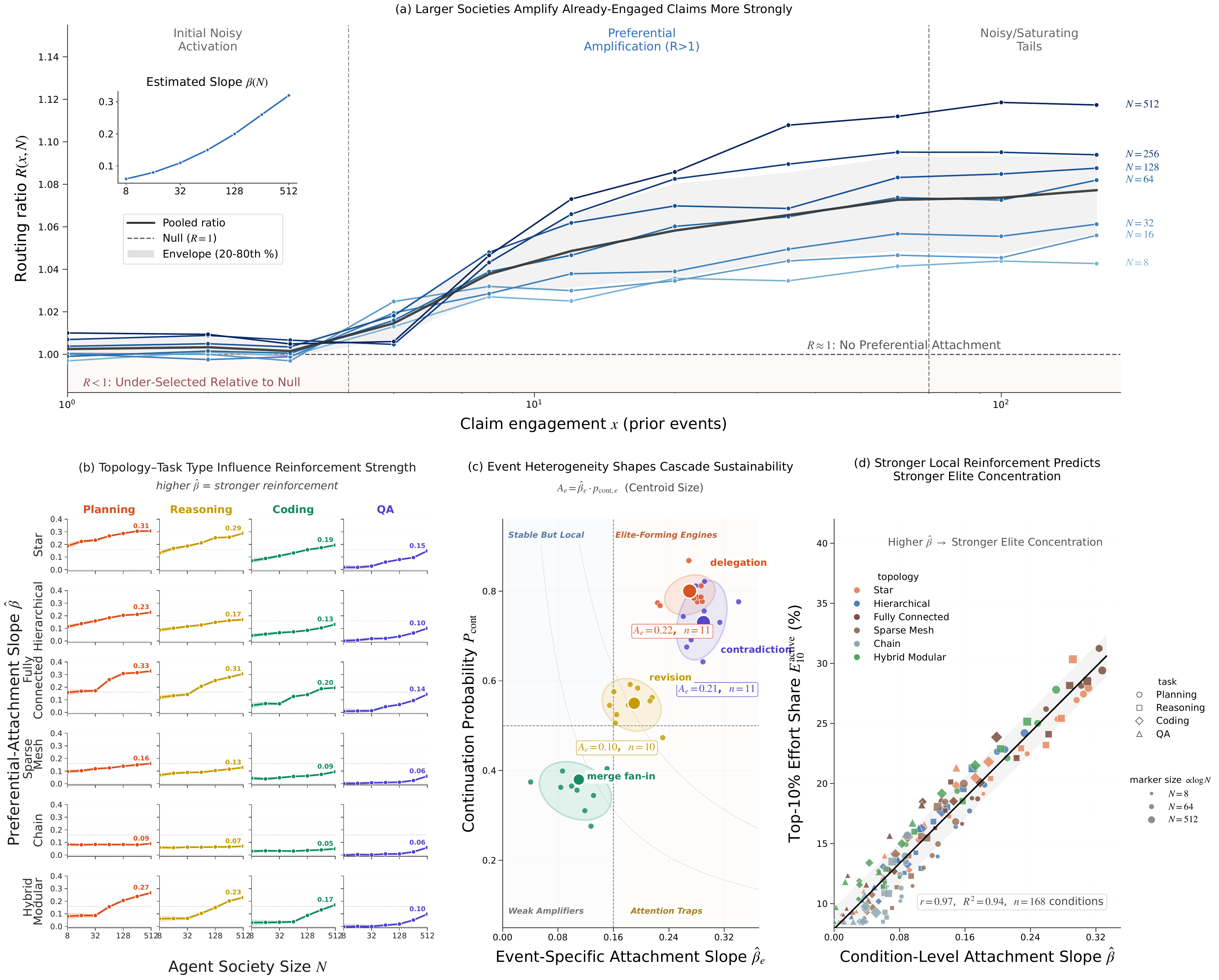}
    \caption{\textbf{Preferential attachment is a core micro-mechanism behind heavy-tailed coordination and elite concentration in LLM agent societies.}
    (a) The routing ratio $R(x,N)$ rises above the null baseline once a claim accumulates prior engagement, and the effect strengthens with system size $N$, revealing scale-dependent preferential amplification before saturating in the tail.
    (b) Estimated attachment slopes $\hat{\beta}$ vary systematically across topology and task type, with stronger reinforcement in star, fully connected, and modular societies than in tree or chain settings.
    (c) Event types differ in cascade-sustaining power: delegation and contradiction occupy the high-amplification, high-continuation region, while revision is intermediate and merge fan-in remains comparatively local.
    (d) Conditions with larger condition-level attachment slope $\hat{\beta}$ also exhibit larger top-10\% effort share $E^{\mathrm{active}}_{10}$, linking local reinforcement directly to macro-level elite concentration.}
    \label{fig:pref-attachment}
    \vspace{-6pt}
\end{figure*}

The mechanism underlying this concentration is \textit{preferential attachment} in claim selection, evidenced directly in Figure~\ref{fig:pref-attachment}. This is consistent with the reinforced routing model in Eq.~\eqref{eq:reinforced_routing} and the attachment-slope definition in Eq.~\eqref{eq:beta_definition}. The routing ratio $R(x,N)$ rises above the null baseline once a claim accumulates prior engagement and strengthens with $N$, implying $\hat{\beta} > 0$. This induces a self-reinforcing dynamic: early-selected claims attract further delegation and contradiction, generating recursive revision loops in which agents repeatedly return to the same active trajectories. Figure~\ref{fig:pref-attachment}(c) shows that delegation and contradiction occupy the high-amplification, high-continuation region of this attachment landscape, while merge fan-in remains local and weakly reinforced. Agents whose claims enter these high-$\hat{\beta}$ loops early remain central over time, giving rise to elite agents. Across conditions, $\hat{\beta}$ predicts $S_{10}$ with $r = 0.97$, linking local reinforcement directly to global concentration.

Figure~\ref{fig:cascade-composition} provides the structural explanation for why this concentration intensifies with scale. As cascade size and $N$ increase, the merge conversion ratio degrades jointly, from 0.21 at small $N$ and short cascades to 0.07 at $N = 512$ in the top-1\% tail. Large cascades are therefore increasingly expansion-heavy and merge-poor: delegation and contradiction compound through recursive revision while integration lags. This integration bottleneck concentrates unresolved coordination onto a shrinking subset of highly engaged agents and their recursive claim loops, amplifying elite dominance as systems scale.

\begin{figure*}[h]
    \centering
    \includegraphics[width=\textwidth]{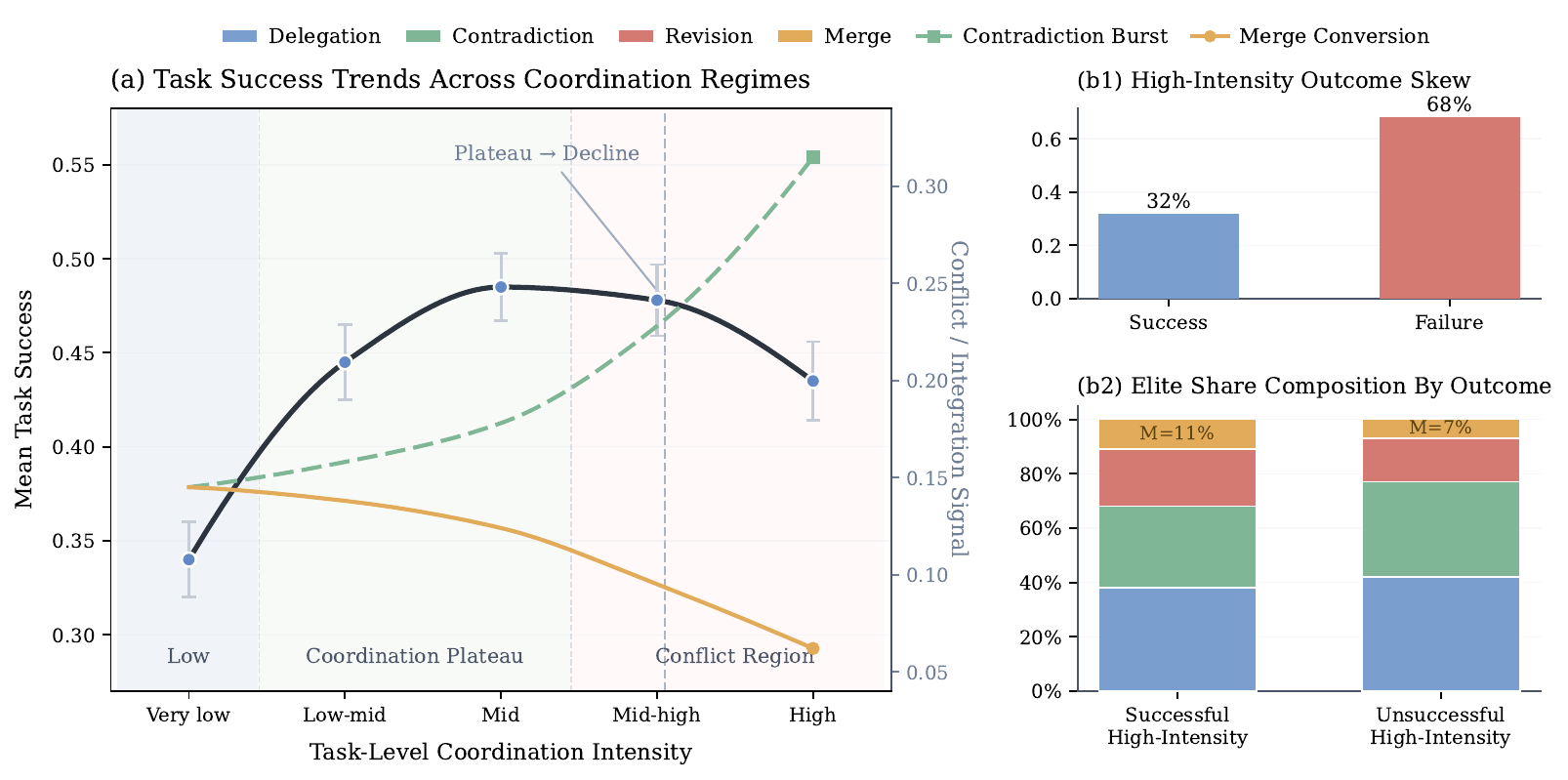}
    \caption{\textbf{Conflict-integration dynamics and performance degradation in the high-intensity regime.} 
    \textbf{(a)} Mean task success across coordination intensity regimes reveals a non-monotonic signature: performance plateaus at moderate intensity before undergoing significant degradation in the high-intensity tail. This decline is strongly correlated with an elevated \textit{contradiction burden} and a concomitant collapse in \textit{merge conversion} efficiency. The vertical dashed line demarcates the transition from productive coordination to the conflict-dominated regime. 
    \textbf{(b1)} Outcome distribution among high-intensity tasks, illustrating a distinct skew toward failure as coordination complexity increases. 
    \textbf{(b2)} Decomposition of elite contribution shares in high-intensity regimes. Successful runs maintain higher integration capacity, whereas unsuccessful runs are characterized by an increased share of contradictory interactions and diminished merge contributions, suggesting that systemic failure is driven by an inability to resolve dense coordination conflicts. $M$ denotes the merge-event share of total elite coordination effort.}
    \label{fig:elite-tail-task-outcome}
    \vspace{-20pt}
\end{figure*}

Figure~\ref{fig:elite-tail-task-outcome} shows the consequence of this structure. Task success peaks at moderate coordination intensity and degrades sharply in the high-intensity tail, where 68\% of runs fail. Failed high-intensity runs exhibit elevated contradiction shares and reduced merge shares relative to successful ones, indicating that elite-dominated, expansion-heavy coordination leads to accumulation without effective consolidation. Across model families (Table~\ref{tab:llm-ablation-global}), $\hat{\beta}$ strongly predicts $E^{\mathrm{active}}_{10}$ ($r = 0.97$)

H2 is supported: intellectual elites emerge endogenously through preferential attachment and recursive revision loops, and their dominance is sustained by an integration bottleneck that intensifies with scale and directly shapes coordination outcomes in large LLM multi-agent systems.

\subsection{H3: Expansion of Extreme Coordination Cascades}

We test whether the magnitude of extreme coordination cascades grows systematically with agent society size by analyzing how $\langle x_{\max} \rangle$ scales with $N$ across coordination primitives.

\begin{figure*}[h]
    \vspace{-3pt}
    \centering
    \includegraphics[width=\textwidth]{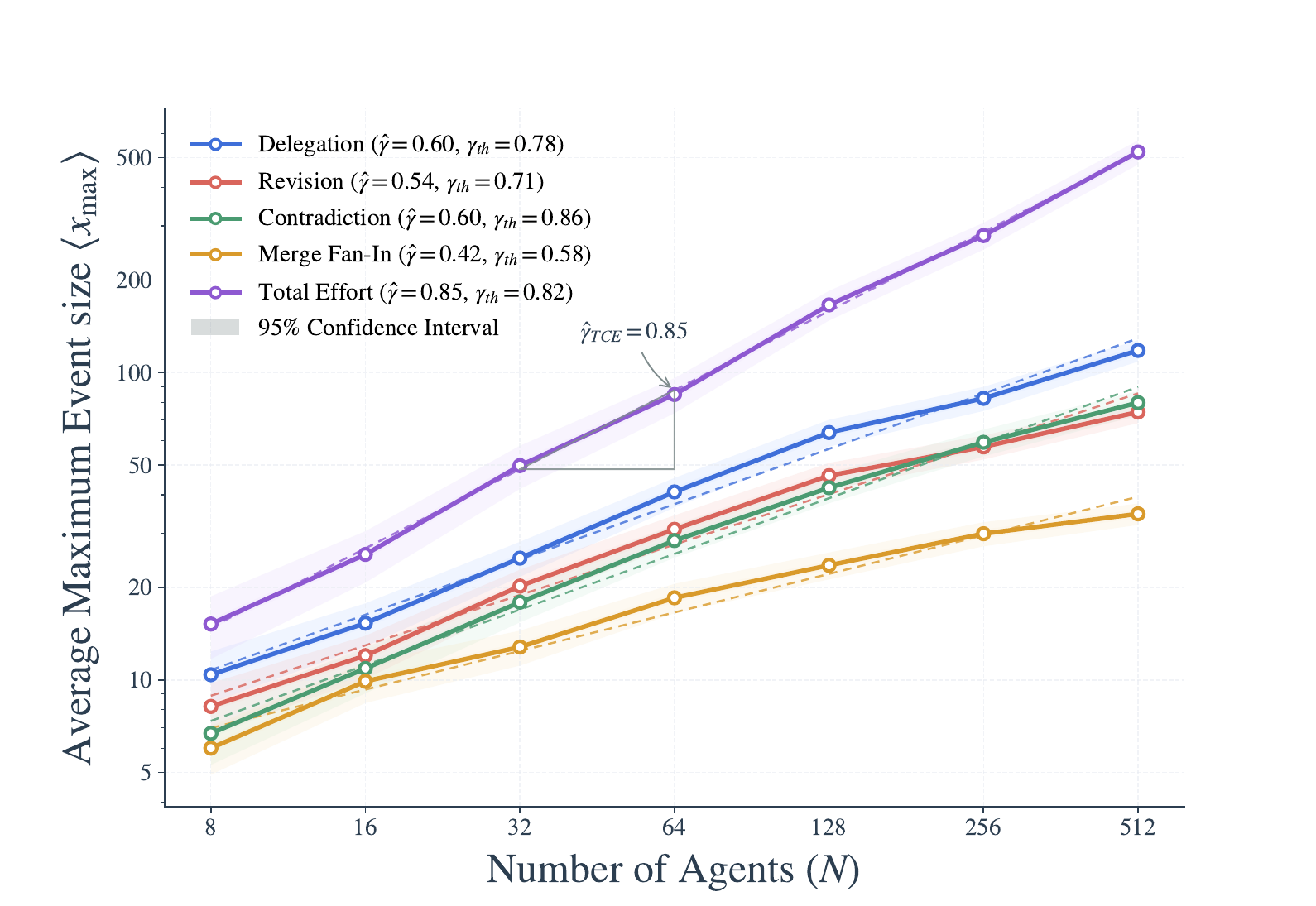}
    \caption{
\textbf{Extreme-value scaling of coordination cascades.}
Mean maximum event size $\langle x_{\max}\rangle$ vs.\ agent count $N$ (log–log). Solid curves show empirical values with 95\% CIs; dashed lines denote power-law fits ($\hat{\gamma}$). All observables scale with $N$, with TCE showing the strongest growth and closest alignment to EVT predictions ($\hat{\gamma}_{\text{TCE}} \approx 0.85$ vs.\ $\gamma_{\text{th}} \approx 0.82$), indicating systematic expansion of the coordination tail.
}
    \label{fig:h3-evt-scaling}
\end{figure*}

Figure~\ref{fig:h3-evt-scaling} provides the primary evidence. $\langle x_{\max} \rangle$ increases systematically with $N$ across all observables on log-log axes, with TCE exhibiting the strongest growth ($\hat{\gamma} \approx 0.85$) and merge fan-in the weakest ($\hat{\gamma} \approx 0.42$). Larger LLM agent societies, therefore, do not merely produce more coordination; they produce qualitatively larger cascades, with the reachable coordination tail expanding by nearly two orders of magnitude from $N=8$ to $N=512$ for TCE.

This scaling is anchored by \textit{Extreme-Value Theory (EVT)}~\cite{embrechts2013modelling, de2006extreme}. Under the truncated power-law form established in H1, the expected maximum scales as $\langle x_{\max} \rangle \propto N^{\gamma}$ with $\gamma_{\mathrm{th}} = 1/(\hat{\alpha}-1)$. The observed TCE exponent $\hat{\gamma} \approx 0.85$ closely matches the theoretical prediction $\gamma_{\mathrm{th}} \approx 0.82$, showing that extreme cascade growth follows directly from the heavy-tailed coordination structure. Other observables fall below their EVT predictions, consistent with finite system constraints suppressing the growth of individual primitives, while TCE, aggregating across all primitives, tracks the theoretical bound most closely.


The divergence in $\hat{\gamma}$ across primitives reflects differences in how coordination processes scale. Delegation and contradiction exhibit stronger scaling because additional agents introduce more opportunities for branching and claim contestation, which compound through preferential attachment and recursive revision loops as $N$ increases. In contrast, merge is integrative and non-generative, lacking a comparable compounding mechanism; its growth is limited to consolidating existing branches rather than creating new ones. As a result, the system becomes increasingly capable of reaching extreme coordination through expansion, while integrative processes scale more slowly, widening the gap between generative and integrative primitives and concentrating extreme cascade mass in the former.

Across model families (Table~\ref{tab:llm-ablation-global}), stronger models reach larger $x_{\max}$ (e.g., GPT-4o-mini at 1320 versus Qwen 2.5 7B at 440), but the scaling structure itself remains invariant. H3 is supported: extreme coordination cascades expand systematically with society size, governed by the heavy-tailed cascade structure of H1 and shaped by the expansion-integration asymmetry of H2.

\textit{H1-H3 reveal a coherent pattern; Coordination in LLM MAS is skewed toward expansion over consolidation, linking scaling failures~\cite{cemri2025multi} to a measurable mechanism and motivating regulation of the expansion-integration balance. We examine this next.}

\section{Law-Aware Intervention: Deficit-Triggered Integration (DTI)}
\label{sec:dti}

\begin{figure*}[t]
    \centering
    \includegraphics[width=\textwidth]{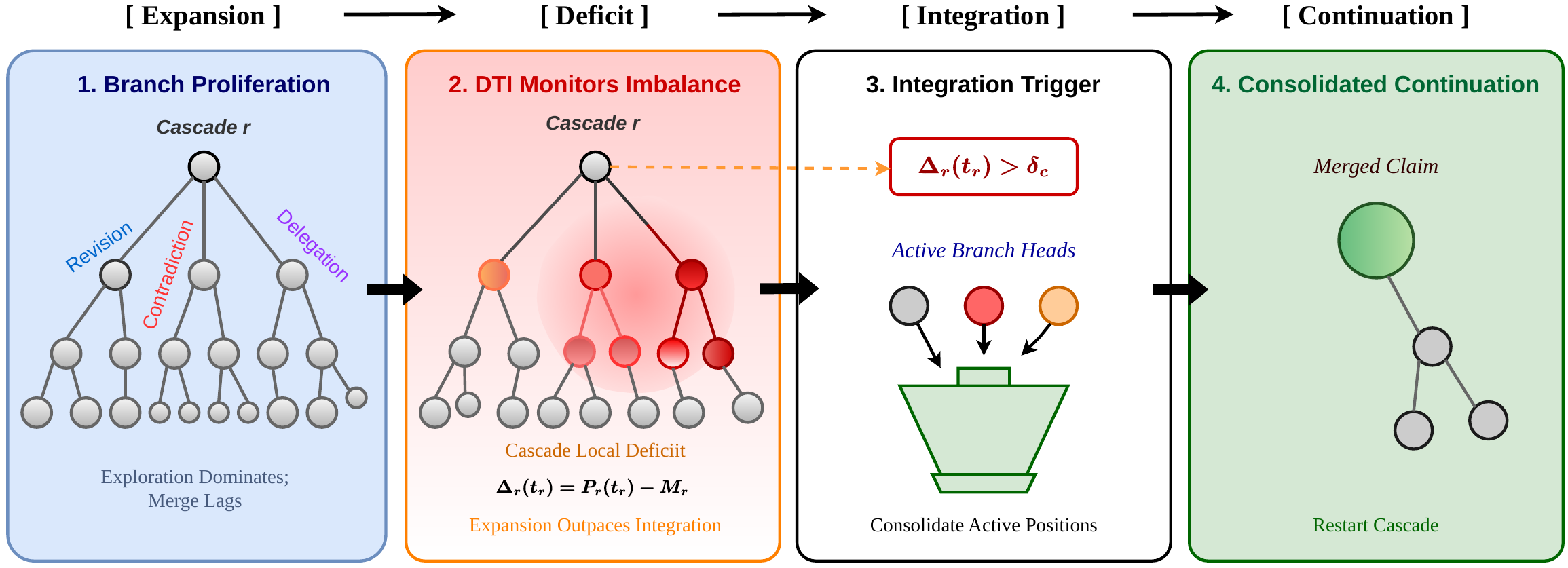}
    \caption{{\textbf{Deficit-Triggered Integration (DTI) in coordination cascades.}
A cascade initially expands through parallel revision, contradiction, and delegation branches, leading to increasing fragmentation (left). DTI monitors this imbalance and triggers integration when it exceeds a threshold, consolidating active branch heads into a unified representation (middle). The cascade then resumes from this integrated state, enabling continued exploration with improved coherence and reduced fragmentation (right).}}
    \label{fig:dti}
    \vspace{-20pt}
\end{figure*}

To directly test whether regulating the expansion–integration balance improves coordination outcomes, we introduce \textit{Deficit-Triggered Integration (DTI)}, a targeted intervention that modifies coordination routing under sustained imbalance. DTI operates at the cascade level by monitoring expansion–integration balance (Appendix~\ref{supp-sec:dti}); when the integration deficit $\Delta_r$ exceeds a condition-specific  threshold $\delta_c$, it temporarily prioritizes integration (Fig.~\ref{fig:dti}). Expansion actions are deferred and agents are routed to merge existing branches, increasing merge fan-in and enforcing consolidation. The intervention is local and state-dependent, activated only under imbalance, and does not alter agent capabilities or impose global constraints.

\begin{figure*}[h]
    \centering
    \includegraphics[width=\textwidth]{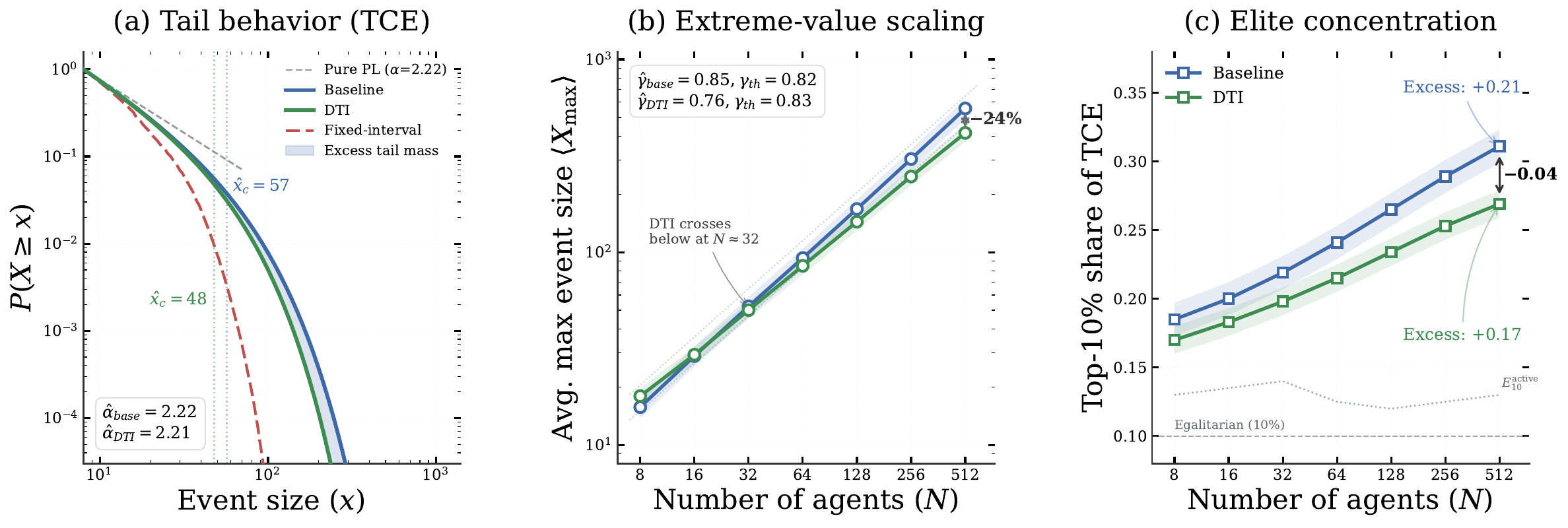}
\caption{\textbf{Impact of Deficit-Triggered Integration (DTI) on collective cognition dynamics.}
(a) DTI preserves the heavy-tailed structure of coordination cascades while shifting truncation earlier, reducing excess tail mass without altering the intermediate scaling regime. Fixed-interval intervention, in contrast, introduces premature truncation and distorts the tail.
(b) The growth of extreme coordination events with system size is preserved under DTI, but large cascades are systematically attenuated, leading to more controlled scaling behavior.
(c) DTI reduces the concentration of cognitive effort among top agents, moderating elite dominance while maintaining a non-uniform and productive coordination structure.
DTI converts late-stage expansion into earlier integration, yielding more stable and balanced collective reasoning without disrupting the underlying heavy-tailed organization of agent coordination.}
    \label{fig:dti-results-main}
\end{figure*}

Figure~\ref{fig:dti-results-main} shows that this reallocation produces measurable structural changes. The merge conversion ratio (merge events as a fraction of  total expansion events per cascade) increases in high-intensity cascades, reversing the degradation observed in Fig.~\ref{fig:cascade-composition}, while contradiction density decreases, indicating improved consolidation of expansion-heavy trajectories. The heavy-tailed distribution of cascade sizes is preserved, showing that large-scale reasoning remains intact while its composition changes. 

\begin{figure*}[t]
    \centering
    \includegraphics[width=\textwidth]{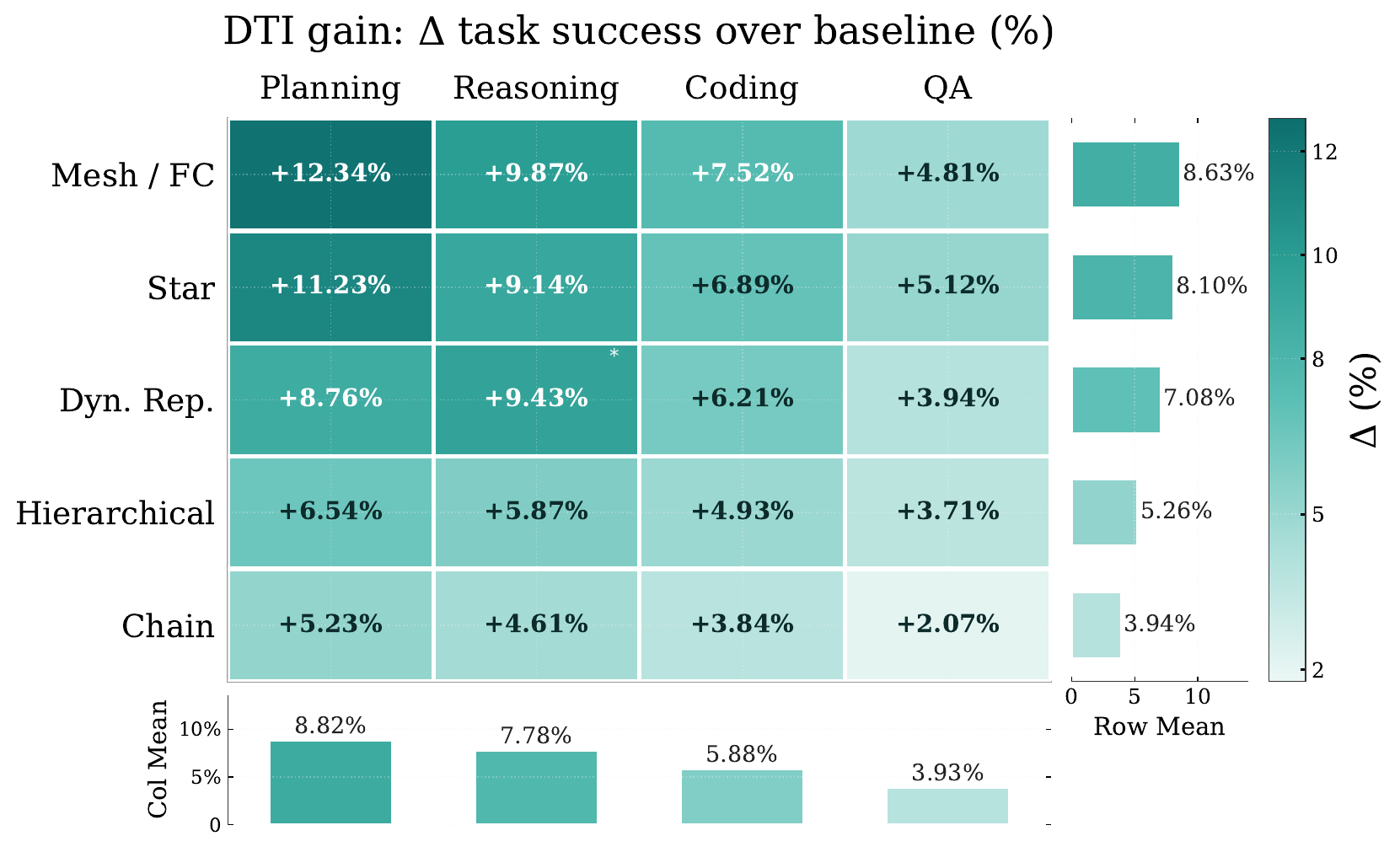}
    \caption{
\textbf{Heterogeneity of DTI gains across topology and task family.} Relative improvement in task success (\%) under DTI versus baseline, reported per topology-task condition. Gains range from 
+2.07\% (QA \(\times\) Chain) to +12.34\% (Planning \(\times\) Mesh/FC). DTI produces the largest 
improvements in conditions exhibiting the strongest 
expansion-integration imbalance in baseline coordination dynamics. Row and column marginals show topology- and task-level averages.}
    \label{fig:dti-accuracy-heatmap}
    \vspace{-8pt}
\end{figure*}

These structural improvements translate into performance gains. Figure~\ref{fig:dti-accuracy-heatmap} shows that gains are largest in high-imbalance settings (Planning$\times$Mesh) and smallest where imbalance is weakest (QA$\times$Chain), confirming that DTI targets the identified failure mode rather than producing uniform improvements. These results establish that the expansion–integration imbalance is causal: selectively increasing integration improves outcomes without suppressing large cascades. DTI thus provides a proof of principle that coordination structure can be directly regulated. Implementation details and the full algorithm are provided in Appendix~\ref{supp-sec:dti}.

\vspace{-5pt}
\section{Discussion and Implications}
\label{sec:discussions}

The central finding of this work is that coordination in LLM multi-agent systems self-organizes into a pattern that is simultaneously productive and self-limiting. The heavy-tailed cascade structure, elite concentration, and systematic growth of extreme events are not independent failure modes; they arise from the same reinforcement dynamics that enable complex collective reasoning. These dynamics amplify coordination while progressively weakening consolidation, leading to non-monotonic returns that cannot be resolved by improving individual agents alone.

This has direct implications for how LLM MAS are designed and evaluated. Current practice treats scaling failures as capability gaps, addressable through better models, prompts, or benchmarks. Our results suggest an additional axis: coordination structure. The tail exponents, attachment slopes, and merge conversion ratios identified here provide a quantitative vocabulary for diagnosing coordination health that is orthogonal to task performance. A system may achieve high accuracy while operating in a structurally fragile configuration, or fail despite sufficient capability if coordination is misallocated. Evaluation frameworks that measure only outcomes miss this dimension.

DTI demonstrates that the coordination structure is regulable, but it is a proof of principle rather than a complete solution. It targets a single imbalance: expansion over consolidation through threshold-triggered integration. A fuller account of coordination regulation would require dynamic topology adaptation, control of reinforcement dynamics, and mechanisms that couple elite formation with effective integration over longer horizons to better distribute utility across agents as the system size increases. The laws developed here provide a quantitative foundation for this broader design space.

\section{Limitations}
\label{sec:limitations}

Our analysis focuses on a controlled set of coordination primitives and topology-task configurations; while the observed patterns are consistent across these settings, additional coordination mechanisms and longer-horizon interactions may introduce dynamics not captured here. Our measurements are derived from discrete event abstractions of coordination, which, while interpretable and comparable across conditions, may not fully capture finer-grained semantic aspects of reasoning quality. DTI is intentionally minimal and targets a single form of imbalance; it does not address all factors influencing coordination outcomes, nor does it guarantee optimal performance across all regimes. Finally, while scaling behavior is consistent across models and settings, our conclusions are empirical and finite-sample in nature, and further theoretical and large-scale validation would strengthen their generality.
\section{Conclusion}
\label{sec:conclusion}

We show that coordination in LLM MAS follows a consistent structural pattern: heavy-tailed cascades, reinforcement-driven concentration, and systematic growth of extreme events. These are coupled effects of how coordination propagates and accumulates, leading, as systems scale, to an increasing imbalance between expansion and integration and to large but weakly consolidated reasoning trajectories. This perspective shifts the focus from agent capability to coordination structure. Scaling failures are not solely due to insufficient reasoning ability, but to how coordination is distributed and reinforced. By identifying measurable laws-tail behavior, attachment dynamics, and integration efficiency, we provide a framework for diagnosing coordination at scale. Our intervention demonstrates that these dynamics are regulable: selectively rebalancing coordination improves outcomes without suppressing large-scale reasoning. Future progress in LLM MAS will therefore depend not only on stronger agents, but on mechanisms that shape how collective reasoning unfolds.


\clearpage

\bibliographystyle{plain} 
\bibliography{main}

\clearpage

\clearpage
\appendix
\section*{Appendix}
\label{supp-sec:appendix}
\renewcommand{\theequation}{A\arabic{equation}}
\setcounter{equation}{0}

\FloatBarrier
\section{Additional Qualitative Results}
\label{app:qualitative-results}

In this section, we provide additional qualitative results pertaining to each of the hypotheses tested:

\begin{figure*}[h]
    \centering
    \includegraphics[width=\textwidth]{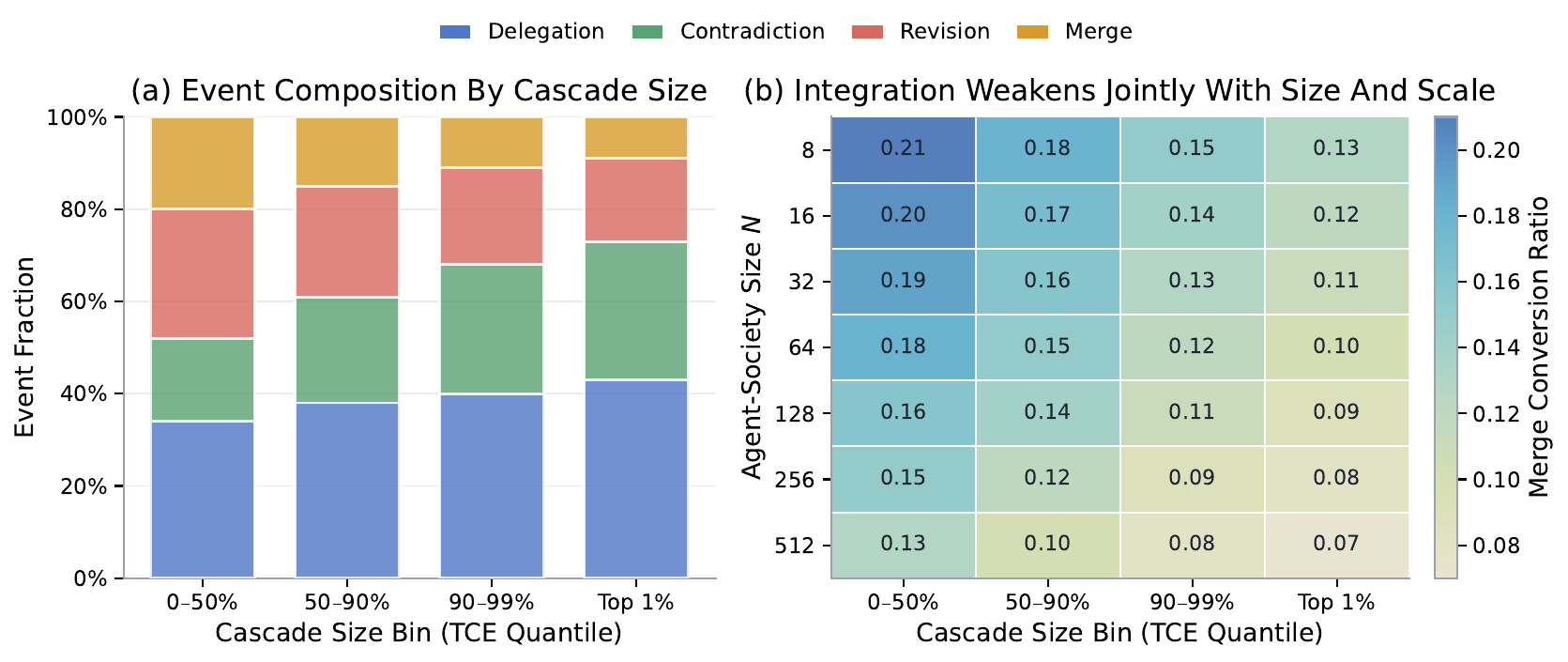}
    \caption{\textbf{Internal composition of claim-level coordination cascades and the scale-conditioned integration bottleneck.}
\textbf{(a)} Claim-rooted cascades are grouped by total cognitive effort (TCE) quantile, pooling all tasks, topologies, and agent scales. As cascades move into the far tail, their internal event composition shifts toward \emph{delegation} and \emph{contradiction}, while \emph{merge} remains comparatively weak and increasingly subproportional; \emph{revision} plays an intermediate role. \textbf{(b)} The same bottleneck appears in the merge conversion ratio across both cascade size and society scale: integration is strongest for smaller cascades in smaller societies and weakens progressively toward extreme cascades in larger societies. Together, these results show that large coordination cascades are increasingly expansion-heavy and merge-poor, and that this integration deficit strengthens with scale.}   
    \label{fig:cascade-composition}
\end{figure*}

\FloatBarrier
\section{Notation}
\label{supp-sec:notations}

We summarize the mathematical notation used throughout the paper.
We use $\beta$ for the theoretical routing strength and $\hat{\beta}$ 
for empirically estimated scaling exponents.
In practice, the theoretical EVT prediction $\gamma_{\mathrm{th}} = 1/(\alpha-1)$ 
is evaluated by substituting the empirical estimate $\hat{\alpha}$.

\subsection{Heavy-Tail Modeling}

\begin{table}[h]
\centering
\small
\begin{tabular}{ll}
\toprule
\textbf{Symbol} & \textbf{Meaning} \\
\midrule
$X$                  & Random variable representing an event-size observable \\
$x$                  & Realized value of $X$ \\
$P(X \ge x)$         & Complementary cumulative distribution function (CCDF) of $X$ \\
$\alpha$             & Tail exponent of the power-law / truncated power-law model \\
$\hat{\alpha}$       & MLE estimate of $\alpha$ \\
$x_{\min}$           & Lower cutoff above which tail fitting is performed \\
$x_c$                & Exponential cutoff scale in the truncated power-law model \\
$\hat{x}_c$          & MLE estimate of $x_c$ \\
$n_{\mathrm{tail}}$  & Number of observations satisfying $x \ge x_{\min}$ \\
$\mathrm{LR}$        & Log-likelihood ratio between two candidate models \\
$p$                  & $p$-value associated with the likelihood-ratio comparison \\
\bottomrule
\end{tabular}
\caption{Notation for heavy-tail modeling and model comparison (H1).}
\end{table}

\subsection{Elite Concentration}

\begin{table}[h]
\centering
\small
\begin{tabular}{ll}
\toprule
\textbf{Symbol} & \textbf{Meaning} \\
\midrule
$\mathrm{TCE}(c_r)$          & Total Cognitive Effort: total number of downstream events from root claim $c_r$ \\
$S_k(c_r)$                   & Top-$k$ contribution share within cascade $\mathcal{C}_r$ \\
$E^{\mathrm{active}}_{k}$    & Top-$k$\% effort share; denominator = active agents only \\
$E^{\mathrm{all}}_{k}$       & Top-$k$\% effort share; denominator = all $N$ agents \\
$\Delta^{\mathrm{active}}_k$ & Excess share above egalitarian baseline $k/N$ \\
$S_p$                        & Cumulative concentration curve (Lorenz-style) \\
$A(N)$                       & Active-agent fraction at society size $N$ \\
$M$                          & Fraction of elite coordination effort attributed to merge events \\
\bottomrule
\end{tabular}
\caption{Notation for elite concentration and inequality (H2).}
\end{table}

\subsection{Extreme-Event Scaling}

\begin{table}[h]
\centering
\small
\begin{tabular}{ll}
\toprule
\textbf{Symbol} & \textbf{Meaning} \\
\midrule
$x_{\max}(N)$                      & Maximum cascade size in a system of $N$ agents \\
$\langle x_{\max} \rangle$         & Expected maximum event size (averaged over runs) \\
$\hat{\gamma}$                     & Empirically estimated extreme-value scaling exponent \\
$\gamma_{\mathrm{th}} = 1/(\alpha-1)$ & Theoretical EVT prediction for scaling exponent \\
\bottomrule
\end{tabular}
\caption{Notation for extreme-event scaling (H3).}
\end{table}

\subsection{Intervention (DTI)}

\begin{table}[h]
\centering
\small
\begin{tabular}{ll}
\toprule
\textbf{Symbol} & \textbf{Meaning} \\
\midrule
$R(x,N)$                              & Routing ratio: relative likelihood a claim with activity $x$ is selected \\
$\beta$                               & Reinforcement strength in the routing rule (Eq.~\ref{eq:beta_definition}) \\
$\hat{\beta}(N)$                      & Empirical condition-level attachment slope \\
$\hat{\beta}_e$                       & Event-specific attachment slope \\
$\hat{\beta}_c$                       & Contradiction scaling exponent used in DTI pressure model \\
$p_{\mathrm{cont},e}$                 & Continuation probability per event type \\
$A_e = \hat{\beta}_e \cdot p_{\mathrm{cont},e}$ & Amplification potential per event type \\
$t_r$                                 & Coordination events elapsed in active cascade segment \\
$M_r$                                 & Realized merge events in current cascade segment \\
$P_r(t_r)$                            & Contradiction-driven exploration pressure (Eq.~\ref{eq:pressure}) \\
$\Delta_r(t_r)$                       & Integration deficit (Eq.~\ref{eq:deficit}) \\
$\delta_c$                            & Condition-specific deficit threshold (Eq.~\ref{eq:trigger}) \\
$a_c$                                 & Normalization constant for condition class $c$ \\
$\mathcal{B}_r$                       & Active branch heads for root claim $r$ \\
\bottomrule
\end{tabular}
\caption{Notation for preferential attachment and Deficit-Triggered Integration (DTI).}
\end{table}

\FloatBarrier
\section{Additional Details on Deficit-Triggered Integration (DTI)}
\label{supp-sec:dti}

We provide a complete formal specification of Deficit-Triggered Integration (DTI), introduced in Section~X. DTI is a cascade-local intervention that monitors the imbalance between exploration and integration and triggers integration when this imbalance exceeds a condition-specific threshold.


DTI operates at the level of individual coordination cascades. Let $r$ denote a root claim, and consider the cascade defined by all events whose causal ancestry traces to $r$. For each active cascade, we maintain a local state consisting of the number of coordination events observed so far, denoted by $t_r$, and the number of realized merge events within the current cascade segment, denoted by $M_r$. These quantities are updated incrementally as events are generated.


To quantify expansion within the cascade, we model contradiction-driven exploration pressure as
\begin{equation}
P_r(t_r) = a_c \, t_r^{\hat{\beta}_{\mathrm{c}}},
\label{eq:pressure}
\end{equation}
where $\hat{\beta}_{\mathrm{c}}$ is the empirically observed scaling exponent for contradiction events, and $a_c$ is a normalization constant defined for each condition class $c$ (topology $\times$ task family).

\begin{algorithm}[h]
\caption{Deficit-Triggered Integration (DTI)}
\label{alg:dti}
\begin{algorithmic}[1]
\Require Event stream $\mathcal{E}=\{e_1,\dots,e_T\}$; contradiction scaling coefficient $\hat{\beta}_{\mathrm{c}}$; condition-specific normalization $a_c$ and threshold $\delta_c$ estimated from baseline logs
\Statex
\State For each active root claim $r$, initialize local cascade state:
\[
t_r \gets 0, \qquad M_r \gets 0
\]
\For{$t = 1$ to $T$}
    \State Observe event $e_t$
    \State Determine its root claim $r \gets \mathrm{root}(e_t)$
    \State Determine its condition class $c \gets \mathrm{cond}(r)$ \Comment{topology $\times$ task family}
    \State Update local cascade length:
    \[
    t_r \gets t_r + 1
    \]

    \If{$\mathrm{type}(e_t)=\textsc{merge}$}
        \State $M_r \gets M_r + 1$
    \EndIf

    \State Compute contradiction-driven exploration pressure:
    \[
    P_r \gets a_c \, t_r^{\hat{\beta}_{\mathrm{c}}}
    \]

    \State Compute integration deficit:
    \[
    \Delta_r \gets P_r - M_r
    \]

    \If{$\Delta_r > \delta_c$}
        \State $\mathcal{B}_r \gets \mathrm{ActiveBranches}(r)$ \Comment{most recent branch-head outputs causally attached to root claim $r$}
        \State Invoke integration over $\mathcal{B}_r$ to produce merged claim $\tilde{e}$
        \State Log $\tilde{e}$ as a \textsc{merge} event attached to root claim $r$
        \State Broadcast $\tilde{e}$ as updated shared context for the cascade rooted at $r$
        \State Update merge count:
        \[
        M_r \gets M_r + 1
        \]
        \State Restart the local cascade segment:
        \[
        t_r \gets 0, \qquad M_r \gets 1
        \]
    \EndIf
\EndFor
\end{algorithmic}
\end{algorithm}

Integration is measured directly through the realized merge count $M_r$. The difference between exploration pressure and realized integration defines the integration deficit:
\begin{equation}
\Delta_r(t_r) = P_r(t_r) - M_r.
\label{eq:deficit}
\end{equation}

DTI triggers an integration step when the deficit exceeds a condition-specific threshold:
\begin{equation}
\Delta_r(t_r) > \delta_c.
\label{eq:trigger}
\end{equation}

At this point, the set of active branch heads is collected as
\begin{equation}
\mathcal{B}_r = \mathrm{ActiveBranches}(r),
\end{equation}
where $\mathcal{B}_r$ consists of the most recent agent outputs whose causal ancestry traces to root claim $r$. A structured integration prompt is applied to $\mathcal{B}_r$, consolidating active positions, identifying agreement, and resolving remaining disagreement into a single merged claim. The resulting output is logged as a \textsc{merge} event and broadcast as updated shared context.

Following integration, the local cascade segment is restarted by setting
\begin{equation}
t_r \gets 0, \qquad M_r \gets 1,
\end{equation}
which reflects that one integration event has already been realized and prevents immediate retriggering.

The parameters $a_c$ and $\delta_c$ are estimated directly from baseline coordination traces for each condition class. The normalization $a_c$ captures the empirical relationship between cascade growth and merge activity, while $\delta_c$ is defined as the mean plus one standard deviation of the integration deficit observed at cascade termination points. These parameters are estimated directly from baseline coordination traces for each condition class and are fixed prior to intervention, introducing no outcome-tuned quantities.

DTI maintains independent state for each active root claim and therefore requires only $O(|R|)$ additional memory, where $|R|$ is the number of active cascades. Each event incurs constant-time updates to $(t_r, M_r)$, and no additional model calls are made except when the deficit threshold is exceeded.

Because both the normalization and threshold are defined per condition class, the frequency of integration adapts automatically to the underlying coordination regime. Cascades with stronger expansion dynamics accumulate deficit more rapidly and therefore trigger integration more frequently, while cascades with weaker imbalance evolve largely unaffected.

\FloatBarrier
\section{Additional Quantitative Results}
\label{app:quantitative}

This section provides additional statistical analyses supporting the empirical
laws presented in the main text. Each subsection addresses a specific aspect
of robustness, inequality, mechanism, or experimental validity.

\subsection{Statistical Validity of Heavy-Tail Fits}
\label{app:fit_robustness}

To assess the robustness of heavy-tail estimates, we report bootstrap confidence
intervals, tail fractions, and goodness-of-fit statistics for each observable.
All observables exhibit stable exponents within the range $\alpha \in [2.1, 2.7]$,
with sufficient tail support and low KS distances, indicating reliable heavy-tail
estimation.

\begin{table}[h]
\centering
\small
\setlength{\tabcolsep}{4pt}
\renewcommand{\arraystretch}{1.1}
\begin{tabular}{lcccc}
\toprule
\textbf{Observable} & $\hat{\alpha}$ & CI & $n_{\text{tail}}/n$ & KS \\
\midrule
Delegation & 2.28 & [2.24, 2.32] & 0.27 & 0.042 \\
Revision & 2.41 & [2.36, 2.46] & 0.31 & 0.047 \\
Contradiction & 2.16 & [2.11, 2.21] & 0.22 & 0.039 \\
Merge & 2.71 & [2.63, 2.79] & 0.14 & 0.051 \\
TCE & 2.22 & [2.18, 2.26] & 0.30 & 0.035 \\
\bottomrule
\end{tabular}
\caption{Robustness of heavy-tail fits across observables. Confidence intervals
are obtained via bootstrap resampling.}
\end{table}

\subsection{Inequality and Elite Concentration}
\label{app:elite_metrics}

Beyond top-$k$ share curves, we quantify inequality using the Gini coefficient
and effective number of agents. As system size increases, both metrics indicate
increasing concentration of reasoning effort, with fewer agents accounting for a
larger fraction of total coordination activity.

\begin{table}[h]
\centering
\small
\setlength{\tabcolsep}{4pt}
\renewcommand{\arraystretch}{1.1}
\begin{tabular}{lcccc}
\toprule
$N$ & Top-1 & Top-10\% & Gini & $N_{\text{eff}}/N$ \\
\midrule
8   & 0.04 & 0.17 & 0.20 & 0.88 \\
32  & 0.07 & 0.21 & 0.25 & 0.81 \\
128 & 0.08 & 0.23 & 0.31 & 0.73 \\
512 & 0.11 & 0.29 & 0.39 & 0.64 \\
\bottomrule
\end{tabular}
\caption{Concentration of reasoning effort across agent scales. Top-$k$ shares
increase while the effective number of contributing agents decreases with $N$.}
\end{table}

\begin{table}[h]
\centering
\small
\begin{tabular}{lcc}
\toprule
Metric & Mean & Std (across seeds) \\
\midrule
Gini & 0.37 & 0.018 \\
Top-10\% share & 0.29 & 0.017 \\
$N_{\text{eff}}/N$ & 0.64 & 0.015 \\
\bottomrule
\end{tabular}
\caption{Stability of inequality metrics across random seeds (averaged over all conditions).}
\end{table}

\subsection{Tail Anatomy of Coordination Cascades}
\label{app:tail_anatomy}

To understand the structure of large coordination events, we analyze the
composition of cascades across size percentiles. Larger cascades are
increasingly dominated by expansion dynamics (delegation and contradiction),
while merge activity declines sharply, indicating a structural imbalance in
integration.

\begin{table}[h]
\centering
\small
\setlength{\tabcolsep}{4pt}
\renewcommand{\arraystretch}{1.1}
\begin{tabular}{lcccc}
\toprule
\textbf{Percentile} & Delegation & Contradiction & Merge & Merge ratio \\
\midrule
Median & 0.30 & 0.21 & 0.19 & 0.37 \\
90th   & 0.36 & 0.26 & 0.15 & 0.24 \\
99th   & 0.41 & 0.31 & 0.11 & 0.15 \\
Top 1\% & 0.45 & 0.34 & 0.08 & 0.10 \\
\bottomrule
\end{tabular}
\caption{Event composition across cascade size percentiles. Expansion dominates
the tail while merge activity diminishes, revealing an integration bottleneck.}
\end{table}

\subsection{Task Expansion Validation}
\label{app:task_expansion}

We report statistics of the benchmark-conditioned task expansion module.
The module generates only workload and dependency structure; coordination
events are not prescribed and are instead extracted from realized execution
traces.

\begin{table}[h]
\centering
\small
\setlength{\tabcolsep}{4pt}
\renewcommand{\arraystretch}{1.1}
\begin{tabular}{lccc}
\toprule
Benchmark & Seeds & Expanded tasks & Avg. dependencies \\
\midrule
GAIA & 30 & 150 & 2.1 \\
SWE-bench & 30 & 150 & 2.4 \\
REALM & 11 & 55 & 2.0 \\
MultiAgentBench & 30 & 150 & 2.3 \\
\bottomrule
\end{tabular}
\caption{Statistics of the task expansion module. Event structure is not injected
but emerges from agent interaction during execution.}
\end{table}

\FloatBarrier
\section{Event-Level Coordination Formulation and Trace Construction}
\label{supp-sec:event-formulation}

\subsection{Coordination Hierarchy}
\label{app:hierarchy}

We distinguish four levels of structure in multi-agent reasoning:

\begin{itemize}
    \item \textbf{Task:} the global problem instance to be solved.
    \item \textbf{Subtask:} a decomposed work unit created via delegation.
    \item \textbf{Claim:} a unit of reasoning, such as a proposed solution, critique, or intermediate result.
    \item \textbf{Event:} a single coordination step that creates, modifies, routes, or combines claims and subtasks.
\end{itemize}

This hierarchy separates \emph{what is being solved} (tasks and subtasks) from \emph{how reasoning evolves} (events acting on claims).
Our analysis operates at the event level: all coordination observables are defined over collections of events rather than over entire tasks.
Importantly, claims represent reasoning artifacts, while events represent transitions between them.

\begin{table}[h]
\centering
\small
\setlength{\tabcolsep}{6pt}
\renewcommand{\arraystretch}{1.2}
\begin{tabular}{ll}
\toprule
\textbf{Claim Type} & \textbf{Description} \\
\midrule
Proposed Claim & An initial statement or solution generated by an agent. \\
Revised Claim & A modification of a prior claim that refines or corrects it. \\
Contradictory Claim & A claim that challenges or disputes an existing claim. \\
Merged Claim & A claim produced by combining multiple parent claims. \\
\bottomrule
\end{tabular}
\caption{
Types of claims observed in coordination traces. Claims represent reasoning artifacts that are created and transformed through events.
}
\label{tab:claim_types}
\end{table}

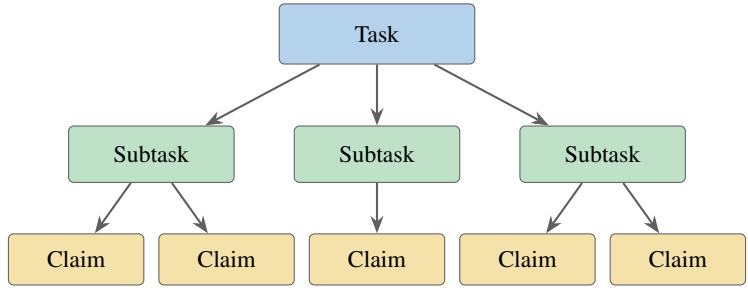
\begin{figure}[h]
\centering
\begin{tikzpicture}[>=Stealth, every node/.style={font=\small}]
    \node[draw=linegray, rounded corners=2pt, fill=pastelblue, minimum width=2.6cm, minimum height=0.8cm] (task) at (0,0) {Task};

    \node[draw=linegray, rounded corners=2pt, fill=pastelgreen, minimum width=2.2cm, minimum height=0.75cm] (s1) at (-3,-1.6) {Subtask};
    \node[draw=linegray, rounded corners=2pt, fill=pastelgreen, minimum width=2.2cm, minimum height=0.75cm] (s2) at (0,-1.6) {Subtask};
    \node[draw=linegray, rounded corners=2pt, fill=pastelgreen, minimum width=2.2cm, minimum height=0.75cm] (s3) at (3,-1.6) {Subtask};

    \node[draw=linegray, rounded corners=2pt, fill=pastelyellow, minimum width=1.8cm, minimum height=0.68cm] (c1) at (-4,-3.0) {Claim};
    \node[draw=linegray, rounded corners=2pt, fill=pastelyellow, minimum width=1.8cm, minimum height=0.68cm] (c2) at (-2,-3.0) {Claim};
    \node[draw=linegray, rounded corners=2pt, fill=pastelyellow, minimum width=1.8cm, minimum height=0.68cm] (c3) at (0,-3.0) {Claim};
    \node[draw=linegray, rounded corners=2pt, fill=pastelyellow, minimum width=1.8cm, minimum height=0.68cm] (c4) at (2,-3.0) {Claim};
    \node[draw=linegray, rounded corners=2pt, fill=pastelyellow, minimum width=1.8cm, minimum height=0.68cm] (c5) at (4,-3.0) {Claim};

    \draw[draw=linegray, line width=0.8pt, ->] (task) -- (s1);
    \draw[draw=linegray, line width=0.8pt, ->] (task) -- (s2);
    \draw[draw=linegray, line width=0.8pt, ->] (task) -- (s3);

    \draw[draw=linegray, line width=0.8pt, ->] (s1) -- (c1);
    \draw[draw=linegray, line width=0.8pt, ->] (s1) -- (c2);
    \draw[draw=linegray, line width=0.8pt, ->] (s2) -- (c3);
    \draw[draw=linegray, line width=0.8pt, ->] (s3) -- (c4);
    \draw[draw=linegray, line width=0.8pt, ->] (s3) -- (c5);

\end{tikzpicture}
\caption{
Hierarchy of coordination structures in a multi-agent system. A task defines the global objective, which is decomposed into subtasks; subtasks produce claims; and claims evolve through event-level coordination steps. Our analysis operates on these events rather than on whole tasks.
}
\label{fig:hierarchy}
\end{figure}

\subsection{Claim and Event Definitions}
\label{app:claim_event_def}

We formalize claims and events as the fundamental elements of coordination dynamics.

\paragraph{Claims:} A claim represents a unit of reasoning produced by an agent, such as a proposed solution, refinement, critique, or synthesis. 
Each claim is assigned a unique identifier and may reference one or more parent claims, forming a directed lineage structure.

Formally, a claim $c_i$ is associated with a set of parent claims $\mathcal{P}(c_i)$, where $|\mathcal{P}(c_i)| \geq 0$. 
A claim with no parents is considered a root claim.

\paragraph{Events:} An event is a coordination action that operates on claims or subtasks. 
Events define how claims are created, modified, and connected, and therefore determine the structure of coordination.

Each event is associated with:
\begin{itemize}
    \item an event type (e.g., \texttt{delegate\_subtask}, \texttt{revise\_claim}, \texttt{contradict\_claim}, \texttt{merge\_claims}),
    \item an acting agent,
    \item a target claim or subtask,
    \item and a resulting claim (if applicable).
\end{itemize}

\paragraph{Event-Induced Claim Transitions:} Each event type induces a specific transformation in the claim structure:

\begin{itemize}
    \item \textbf{Revision:} creates a new claim with a single parent, forming a chain.
    \item \textbf{Contradiction:} creates a new claim that challenges an existing claim, introducing branching.
    \item \textbf{Merge:} creates a new claim with multiple parent claims, introducing multi-parent dependencies.
    \item \textbf{Delegation:} creates a new subtask and initiates new claims within that subtask.
\end{itemize}

These transitions define the edges of the claim graph and collectively produce a directed acyclic graph (DAG) over claims.

\begin{table}[t]
\centering
\small
\setlength{\tabcolsep}{6pt}
\renewcommand{\arraystretch}{1.2}
\begin{tabular}{lll}
\toprule
\textbf{Event Type} & \textbf{Operation on Claims} & \textbf{Resulting Structure} \\
\midrule
\texttt{revise\_claim} & single parent $\rightarrow$ child & chain \\
\texttt{contradict\_claim} & parent $\rightarrow$ multiple children & branching \\
\texttt{merge\_claims} & multiple parents $\rightarrow$ child & multi-parent (DAG) \\
\texttt{delegate\_subtask} & creates new subtask context & hierarchical tree \\
\bottomrule
\end{tabular}
\caption{
Event types and their induced transformations on the claim structure. These operations define how the claim graph evolves from chains to branching structures and ultimately to a directed acyclic graph through merge operations.
}
\label{tab:event_transformations}
\end{table}

\paragraph{Key Property:} The coordination structure is not imposed a priori but emerges from event-induced claim transitions. 
In particular, merge operations introduce multi-parent dependencies, converting tree-like reasoning structures into directed acyclic graphs.

\begin{figure}[t]
\centering
\begin{tikzpicture}[>=Stealth, every node/.style={font=\small}]

\node[draw=linegray, circle, fill=pastelyellow, minimum size=6mm] (r1) at (0,0) {};
\node[draw=linegray, circle, fill=pastelyellow, minimum size=6mm] (r2) at (1.2,0) {};
\node[draw=linegray, circle, fill=pastelyellow, minimum size=6mm] (r3) at (2.4,0) {};

\draw[draw=linegray, ->] (r1) -- (r2);
\draw[draw=linegray, ->] (r2) -- (r3);

\node at (1.2,-0.9) {\footnotesize Revision (Chain)};

\node[draw=linegray, circle, fill=pastelyellow, minimum size=6mm] (c1) at (5,0) {};
\node[draw=linegray, circle, fill=pastelpink, minimum size=6mm] (c2) at (4.2,-1.2) {};
\node[draw=linegray, circle, fill=pastelpink, minimum size=6mm] (c3) at (5.8,-1.2) {};

\draw[draw=linegray, ->] (c2) -- (c1);
\draw[draw=linegray, ->] (c3) -- (c1);

\node at (5,-2.2) {\footnotesize Contradiction (Branching)};

\node[draw=linegray, circle, fill=pastelyellow, minimum size=6mm] (m1) at (9,-1.2) {};
\node[draw=linegray, circle, fill=pastelyellow, minimum size=6mm] (m2) at (10.4,-1.2) {};
\node[draw=linegray, circle, fill=pastelpurple, minimum size=6mm] (m3) at (9.7,0) {};

\draw[draw=linegray, ->] (m1) -- (m3);
\draw[draw=linegray, ->] (m2) -- (m3);

\node at (9.7,-2.2) {\footnotesize Merge (Multi-parent)};

\end{tikzpicture}

\caption{
Event-induced transformations of the claim structure. 
Revision produces linear chains, contradiction introduces branching around a claim, and merge combines multiple parent claims into a single node, yielding a directed acyclic graph.
}
\label{fig:event_transformations}
\end{figure}
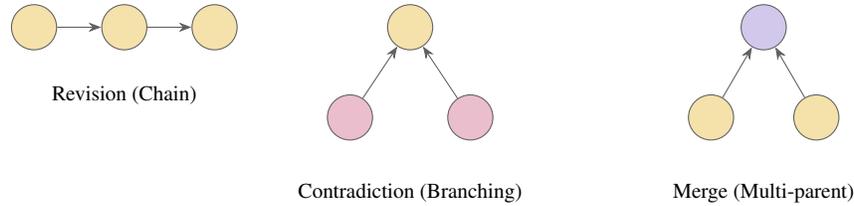

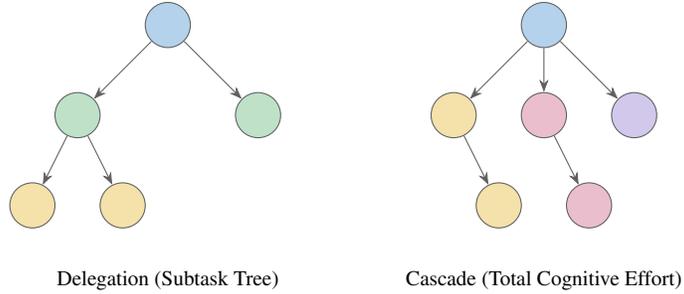
\begin{figure}[h]
\centering
\begin{tikzpicture}[>=Stealth, every node/.style={font=\small}]

\node[draw=linegray, circle, fill=pastelblue, minimum size=6mm] (t0) at (0,0) {};
\node[draw=linegray, circle, fill=pastelgreen, minimum size=6mm] (t1) at (-1.2,-1.2) {};
\node[draw=linegray, circle, fill=pastelgreen, minimum size=6mm] (t2) at (1.2,-1.2) {};
\node[draw=linegray, circle, fill=pastelyellow, minimum size=6mm] (t3) at (-1.8,-2.4) {};
\node[draw=linegray, circle, fill=pastelyellow, minimum size=6mm] (t4) at (-0.6,-2.4) {};

\draw[draw=linegray, ->] (t0) -- (t1);
\draw[draw=linegray, ->] (t0) -- (t2);
\draw[draw=linegray, ->] (t1) -- (t3);
\draw[draw=linegray, ->] (t1) -- (t4);

\node at (0,-3.4) {\footnotesize Delegation (Subtask Tree)};

\node[draw=linegray, circle, fill=pastelblue, minimum size=6mm] (c0) at (5,0) {};
\node[draw=linegray, circle, fill=pastelyellow, minimum size=6mm] (c1) at (3.8,-1.2) {};
\node[draw=linegray, circle, fill=pastelpink, minimum size=6mm] (c2) at (5,-1.2) {};
\node[draw=linegray, circle, fill=pastelpurple, minimum size=6mm] (c3) at (6.2,-1.2) {};
\node[draw=linegray, circle, fill=pastelyellow, minimum size=6mm] (c4) at (4.4,-2.4) {};
\node[draw=linegray, circle, fill=pastelpink, minimum size=6mm] (c5) at (5.6,-2.4) {};

\draw[draw=linegray, ->] (c0) -- (c1);
\draw[draw=linegray, ->] (c0) -- (c2);
\draw[draw=linegray, ->] (c0) -- (c3);
\draw[draw=linegray, ->] (c1) -- (c4);
\draw[draw=linegray, ->] (c2) -- (c5);

\node at (5,-3.4) {\footnotesize Cascade (Total Cognitive Effort)};

\end{tikzpicture}

\caption{
Delegation and cascade structure. 
Delegation events construct a subtask tree (left), defining how work is decomposed. 
Coordination cascades (right) capture the total downstream activity triggered by a root claim, combining revision, contradiction, and merge dynamics. 
Total Cognitive Effort (TCE) is defined as the size of this reachable cascade.
}
\label{fig:delegation_tce}
\end{figure}

\subsection{Trace Logging Schema}
\label{app:trace_schema}

We record fine-grained coordination traces at the event level to enable reconstruction of coordination structures. 
Each interaction step produces a structured record containing event metadata, claim lineage, subtask context, and derived coordination attributes.

\paragraph{Overview:} Each logged entry corresponds to a single coordination event and includes identifiers that link it to claims, subtasks, and other events. 
These records collectively define the inputs required to reconstruct both the subtask tree and the claim DAG.

\paragraph{Event-Level Fields.}
\begin{table}[t]
\centering
\small
\setlength{\tabcolsep}{6pt}
\renewcommand{\arraystretch}{1.2}
\begin{tabular}{ll}
\toprule
\textbf{Field} & \textbf{Description} \\
\midrule
run\_id & Identifier for the experiment run \\
step\_id & Temporal ordering of events \\
agent\_id & Acting agent \\
event\_type & Type of coordination event \\
target\_claim\_id & Referenced claim (if applicable) \\
target\_subtask\_id & Referenced subtask (if applicable) \\
timestamp & Event time \\
message\_length & Token count (proxy for cost) \\
\bottomrule
\end{tabular}
\caption{
Event-level fields recorded for each coordination step.
}
\label{tab:event_fields}
\end{table}

\paragraph{Claim Graph Fields:}
\begin{table}[t]
\centering
\small
\setlength{\tabcolsep}{6pt}
\renewcommand{\arraystretch}{1.2}
\begin{tabular}{ll}
\toprule
\textbf{Field} & \textbf{Description} \\
\midrule
claim\_id & Unique identifier for each claim \\
parent\_claim\_ids & Parent claims defining DAG edges \\
root\_claim\_id & Root ancestor of the claim \\
claim\_depth & Depth within the claim DAG \\
claim\_status & Proposed, revised, contradictory, or merged \\
\bottomrule
\end{tabular}
\caption{
Fields defining the claim-level DAG structure.
}
\label{tab:claim_fields}
\end{table}

The \texttt{root\_claim\_id} is propagated across lineage and enables reconstruction of full coordination cascades.

\paragraph{Subtask Tree Fields.}
\begin{table}[t]
\centering
\small
\setlength{\tabcolsep}{6pt}
\renewcommand{\arraystretch}{1.2}
\begin{tabular}{ll}
\toprule
\textbf{Field} & \textbf{Description} \\
\midrule
subtask\_id & Unique identifier for each subtask \\
parent\_subtask\_id & Parent subtask in hierarchy \\
subtask\_depth & Depth in subtask tree \\
assigned\_agent & Responsible agent \\
subtask\_status & Active or completed \\
\bottomrule
\end{tabular}
\caption{
Fields defining the hierarchical subtask decomposition.
}
\label{tab:subtask_fields}
\end{table}

\paragraph{Derived Coordination Fields.}
\begin{table}[t]
\centering
\small
\setlength{\tabcolsep}{6pt}
\renewcommand{\arraystretch}{1.2}
\begin{tabular}{ll}
\toprule
\textbf{Field} & \textbf{Description} \\
\midrule
revision\_chain\_id & Groups claims in a revision sequence \\
contradiction\_group\_id & Groups claims targeting the same parent claim \\
merge\_id & Identifier for merge operations \\
merge\_parent\_ids & Parent claims involved in merge \\
\bottomrule
\end{tabular}
\caption{
Derived fields used to group events into coordination structures.
}
\label{tab:derived_fields}
\end{table}

These fields are computed from raw interaction traces and are not directly provided by the system. 
They enable identification of revision waves, contradiction bursts, and merge operations during post-processing.

\paragraph{Reconstructability:} Together, these fields provide sufficient information to reconstruct:

\begin{itemize}
    \item the subtask tree from delegation events,
    \item the claim DAG from parent--child relationships,
    \item and coordination cascades from root claim identifiers.
\end{itemize}

\begin{figure*}[t]
\centering
\begin{tikzpicture}[>=Stealth, font=\small, scale=0.92, transform shape]

\tikzset{
  panel/.style={
    draw=linegray,
    rounded corners=2pt,
    minimum width=3.9cm,
    minimum height=4.7cm
  },
  tracebox/.style={
    draw=linegray,
    rounded corners=2pt,
    fill=pastelgreen,
    minimum width=2.95cm,
    minimum height=0.95cm,
    align=left,
    inner sep=3pt,
    font=\scriptsize
  },
  claimnode/.style={
    circle,
    draw=linegray,
    line width=0.8pt,
    minimum size=5.8mm,
    inner sep=0pt
  },
  pipe/.style={
    draw=linegray,
    line width=0.85pt,
    -{Stealth[length=1.8mm,width=1.2mm]}
  }
}

\node[panel] (L) at (0,0) {};
\node[font=\bfseries] at (0,1.85) {Event Traces};

\node[tracebox] (log1) at (0,0.95) {\texttt{revise\_claim}\\
\texttt{claim\_id: c2}\\
\texttt{parent: [c1]}};

\node[tracebox] (log2) at (0,-0.15) {\texttt{contradict\_claim}\\
\texttt{claim\_id: c3}\\
\texttt{parent: [c1]}};

\node[tracebox] (log3) at (0,-1.25) {\texttt{merge\_claims}\\
\texttt{claim\_id: c5}\\
\texttt{parents: [c2,c4]}};

\node[panel] (M) at (5.0,0) {};
\node[font=\bfseries] at (5.0,1.85) {Claim DAG};

\node[claimnode, fill=pastelblue]   (c1) at (5.0,1.0) {};
\node[claimnode, fill=pastelyellow] (c2) at (4.1,0.0) {};
\node[claimnode, fill=pastelpink]   (c3) at (5.9,0.0) {};
\node[claimnode, fill=pastelyellow] (c4) at (4.1,-1.1) {};
\node[claimnode, fill=pastelpurple] (c5) at (5.0,-2.0) {};

\node[font=\scriptsize] at (4.72,1.38) {$c_1$};
\node[font=\scriptsize] at (3.72,-0.02) {$c_2$};
\node[font=\scriptsize] at (6.28,-0.02) {$c_3$};
\node[font=\scriptsize] at (3.72,-1.12) {$c_4$};
\node[font=\scriptsize] at (5.0,-2.42) {$c_5$};

\draw[pipe] (c1) -- (c2);
\draw[pipe] (c1) -- (c3);
\draw[pipe] (c2) -- (c4);
\draw[pipe] (c2) -- (c5);
\draw[pipe] (c4) -- (c5);

\node[panel] (R) at (10.0,0) {};
\node[font=\bfseries] at (10.0,1.85) {Cascade};

\draw[draw=linegray, dashed, rounded corners=4pt, line width=0.8pt]
    (8.7,1.45) rectangle (11.3,-2.35);

\node[claimnode, fill=pastelblue]   (d1) at (10.0,1.0) {};
\node[claimnode, fill=pastelyellow] (d2) at (9.1,0.0) {};
\node[claimnode, fill=pastelpink]   (d3) at (10.9,0.0) {};
\node[claimnode, fill=pastelyellow] (d4) at (9.1,-1.1) {};
\node[claimnode, fill=pastelpurple] (d5) at (10.0,-2.0) {};

\draw[pipe] (d1) -- (d2);
\draw[pipe] (d1) -- (d3);
\draw[pipe] (d2) -- (d4);
\draw[pipe] (d2) -- (d5);
\draw[pipe] (d4) -- (d5);

\draw[pipe] (1.95,0) -- (3.05,0);
\draw[pipe] (6.95,0) -- (8.05,0);

\end{tikzpicture}
\caption{
From structured event traces to coordination cascades. Logged event fields define parent--child claim relationships, from which we reconstruct the claim DAG. Cascades are then extracted as root-centered reachable subgraphs, enabling computation of event-level observables such as revision waves, contradiction bursts, merge fan-in, and total cognitive effort.
}
\label{fig:logs_to_dag_to_cascade}
\end{figure*}
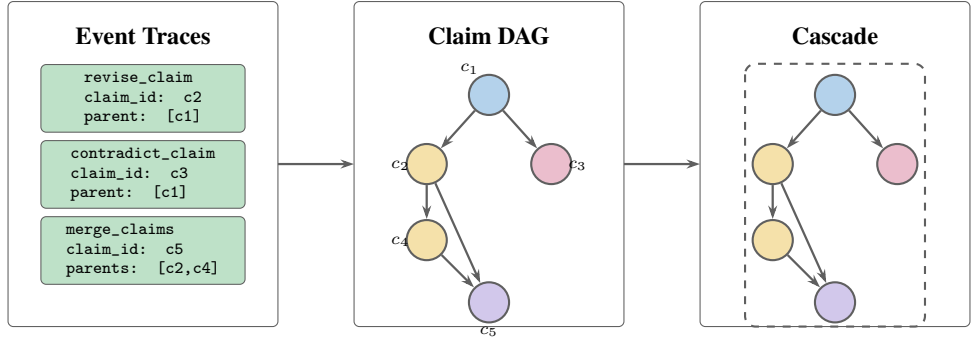

\subsection{DAG Construction Pipeline}
\label{app:dag_construction}

We reconstruct coordination structures from event traces in two stages: (i) subtask-tree construction from delegation events, and (ii) claim-DAG construction from claim-level lineage fields. Figure~\ref{fig:logs_to_dag_to_cascade} summarizes this pipeline.

\paragraph{Subtask Tree Construction.}
Delegation events define a hierarchical decomposition over work units. 
For each \texttt{delegate\_subtask} event, we create a new subtask node and add a directed edge from \texttt{parent\_subtask\_id} to \texttt{subtask\_id}. 
This produces a rooted tree over subtasks, where depth corresponds to recursive decomposition depth. 
The subtask tree therefore captures \emph{what agents are asked to work on}.

\paragraph{Claim DAG Construction.}
Claims form a separate structure that captures \emph{how reasoning evolves}. 
Each claim is represented as a node indexed by \texttt{claim\_id}. 
Directed edges are created from every element of \texttt{parent\_claim\_ids} to the current claim. 
This yields three canonical cases:

\begin{itemize}
    \item \textbf{Revision:} a claim has exactly one parent, producing a chain.
    \item \textbf{Contradiction:} multiple child claims may reference the same parent, producing branching.
    \item \textbf{Merge:} a claim has multiple parents, producing a multi-parent node.
\end{itemize}

Because merge operations introduce multiple incoming edges, the resulting structure is in general a directed acyclic graph (DAG), rather than a tree.

\paragraph{Root Claim Assignment.}
Each claim is associated with a \texttt{root\_claim\_id}, inherited from its earliest ancestor. 
Claims with no parents are initialized as root claims. 
For descendant claims, the root identifier is propagated through lineage during post-processing. 
This field is critical because it makes cascades reconstructable from local parent--child relationships.

\paragraph{Derived Coordination Grouping.}
Several higher-level coordination structures are not directly emitted by the system and must be derived from raw traces:

\begin{itemize}
    \item \textbf{Revision chains} are formed by grouping claims linked through iterative single-parent revision steps.
    \item \textbf{Contradiction groups} are formed by grouping claims that reference the same parent claim within a temporal window.
    \item \textbf{Merge groups} are formed by collecting all parent claims participating in a \texttt{merge\_claims} event.
\end{itemize}

These derived identifiers are used to extract revision waves, contradiction bursts, and merge fan-in during analysis. The trace schema therefore records both raw lineage fields and post-processed grouping fields needed for event-level measurement. 

\paragraph{Cascade Extraction.}
Given a root claim $c_{\mathrm{root}}$, we define its coordination cascade as the reachable subgraph of all downstream claim and event instances associated with that root. 
Operationally, cascade extraction begins from \texttt{root\_claim\_id} and traverses all descendant claims connected through revision, contradiction, merge, and delegation-linked activity. 
This reachable subgraph is the object on which aggregate observables, such as total cognitive effort, are defined.

\paragraph{Subtask Tree vs.\ Claim DAG.}
The subtask tree and claim DAG serve different purposes and should not be conflated. 
The subtask tree records task decomposition induced by delegation, while the claim DAG records the emergent propagation and integration of reasoning. 
In particular, the subtask tree prescribes work structure, whereas the claim DAG captures coordination structure. 
Our event-level observables are defined primarily on the claim DAG and its associated cascades, with delegation cascades obtained from the subtask tree.

\paragraph{Reproducibility.}
This construction procedure ensures that all coordination observables reported in the paper are computed from structured traces rather than manually annotated or heuristically imposed interaction graphs. 
Given the logged fields in Section~\ref{app:trace_schema}, the subtask tree, claim DAG, and cascade assignments are fully reconstructable.

\subsection{Worked Example: From Task Expansion to Coordination Cascade}
\label{app:worked_example}

We illustrate the full coordination pipeline using a representative configuration from our experimental setup. 
We consider a system with $N=16$ agents operating on a GAIA-style multi-step reasoning task. 
As described in Section~3, each task is expanded into an interdependent task tree using a structured reasoning procedure.

\paragraph{Task Expansion.}
Given an input task requiring multi-step reasoning and constraint satisfaction, the expansion module generates the following interdependent task structure:

\begin{itemize}
    \item \textbf{T0:} Solve the full problem
    \item \textbf{T1:} Extract key entities and constraints from the prompt
    \item \textbf{T2:} Generate candidate reasoning paths
    \item \textbf{T3:} Evaluate candidates under extracted constraints
    \item \textbf{T4:} Refine top candidates
    \item \textbf{T5:} Integrate reasoning into final answer
\end{itemize}

Dependencies are not purely hierarchical: T3 depends on both T1 and T2, T4 depends on T3, and T5 integrates outputs from T2 and T4. 
This produces a directed task graph (referred to as a task tree for simplicity) with cross-subtask dependencies.

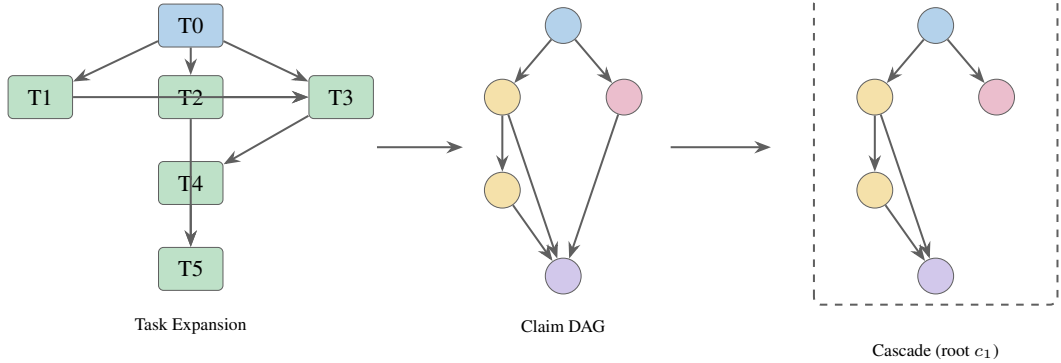
\begin{figure*}[t]
\centering
\resizebox{\textwidth}{!}{%
\begin{tikzpicture}[>=Stealth, font=\small, transform shape]

\node[draw=linegray, rounded corners=2pt, fill=pastelblue, minimum width=0.9cm, minimum height=0.6cm] (t0) at (0,1.7) {T0};

\node[draw=linegray, rounded corners=2pt, fill=pastelgreen, minimum width=0.9cm, minimum height=0.6cm] (t1) at (-2.1,0.7) {T1};
\node[draw=linegray, rounded corners=2pt, fill=pastelgreen, minimum width=0.9cm, minimum height=0.6cm] (t2) at (0,0.7) {T2};
\node[draw=linegray, rounded corners=2pt, fill=pastelgreen, minimum width=0.9cm, minimum height=0.6cm] (t3) at (2.1,0.7) {T3};

\node[draw=linegray, rounded corners=2pt, fill=pastelgreen, minimum width=0.9cm, minimum height=0.6cm] (t4) at (0,-0.5) {T4};
\node[draw=linegray, rounded corners=2pt, fill=pastelgreen, minimum width=0.9cm, minimum height=0.6cm] (t5) at (0,-1.7) {T5};

\draw[->, draw=linegray, line width=0.8pt] (t0) -- (t1);
\draw[->, draw=linegray, line width=0.8pt] (t0) -- (t2);
\draw[->, draw=linegray, line width=0.8pt] (t0) -- (t3);

\draw[->, draw=linegray, line width=0.8pt] (t1) -- (t3);
\draw[->, draw=linegray, line width=0.8pt] (t2) -- (t3);
\draw[->, draw=linegray, line width=0.8pt] (t3) -- (t4);
\draw[->, draw=linegray, line width=0.8pt] (t2) -- (t5);
\draw[->, draw=linegray, line width=0.8pt] (t4) -- (t5);

\node at (0,-2.5) {\scriptsize Task Expansion};

\node[circle, draw=linegray, fill=pastelblue, minimum size=5mm, inner sep=0pt] (c1) at (5.2,1.7) {};
\node[circle, draw=linegray, fill=pastelyellow, minimum size=5mm, inner sep=0pt] (c2) at (4.35,0.7) {};
\node[circle, draw=linegray, fill=pastelpink, minimum size=5mm, inner sep=0pt] (c3) at (6.05,0.7) {};
\node[circle, draw=linegray, fill=pastelyellow, minimum size=5mm, inner sep=0pt] (c4) at (4.35,-0.6) {};
\node[circle, draw=linegray, fill=pastelpurple, minimum size=5mm, inner sep=0pt] (c5) at (5.2,-1.8) {};

\draw[->, draw=linegray, line width=0.8pt] (c1) -- (c2);
\draw[->, draw=linegray, line width=0.8pt] (c1) -- (c3);
\draw[->, draw=linegray, line width=0.8pt] (c2) -- (c4);
\draw[->, draw=linegray, line width=0.8pt] (c2) -- (c5);
\draw[->, draw=linegray, line width=0.8pt] (c4) -- (c5);
\draw[->, draw=linegray, line width=0.8pt] (c3) -- (c5);

\node at (5.2,-2.5) {\scriptsize Claim DAG};

\node[circle, draw=linegray, fill=pastelblue, minimum size=5mm, inner sep=0pt] (d1) at (10.4,1.7) {};
\node[circle, draw=linegray, fill=pastelyellow, minimum size=5mm, inner sep=0pt] (d2) at (9.55,0.7) {};
\node[circle, draw=linegray, fill=pastelpink, minimum size=5mm, inner sep=0pt] (d3) at (11.25,0.7) {};
\node[circle, draw=linegray, fill=pastelyellow, minimum size=5mm, inner sep=0pt] (d4) at (9.55,-0.6) {};
\node[circle, draw=linegray, fill=pastelpurple, minimum size=5mm, inner sep=0pt] (d5) at (10.4,-1.8) {};

\draw[->, draw=linegray, line width=0.8pt] (d1) -- (d2);
\draw[->, draw=linegray, line width=0.8pt] (d1) -- (d3);
\draw[->, draw=linegray, line width=0.8pt] (d2) -- (d4);
\draw[->, draw=linegray, line width=0.8pt] (d2) -- (d5);
\draw[->, draw=linegray, line width=0.8pt] (d4) -- (d5);

\draw[dashed, draw=linegray, rounded corners=3pt, line width=0.8pt] (8.7,-2.2) rectangle (12.1,2.1);

\node at (10.4,-2.85) {\scriptsize Cascade (root $c_1$)};

\draw[->, draw=linegray, line width=0.8pt] (2.6,0) -- (3.8,0);
\draw[->, draw=linegray, line width=0.8pt] (6.7,0) -- (8.1,0);

\end{tikzpicture}%
}
\caption{
End-to-end coordination pipeline in our experimental setup. 
Left: task expansion produces an interdependent task tree with cross-subtask dependencies. 
Center: agent interactions generate claims and event-induced edges, forming a claim DAG. 
Right: the cascade extracted from the root claim captures the full coordination effort used to compute event-level observables.
}
\label{fig:worked_example_pipeline}
\end{figure*}

\paragraph{Task Tree.}
Figure~\ref{fig:worked_example_pipeline} (left) shows the resulting task structure.

\paragraph{Agent Execution and Event Trace.}
Agents are assigned to subtasks and generate claims through interaction. 
A representative trace excerpt is:

\begin{itemize}
    \item \texttt{c1: propose\_claim (T2)} — initial reasoning path
    \item \texttt{c2: revise\_claim(c1) (T3)} — adjusted using constraints from T1
    \item \texttt{c3: contradict\_claim(c1) (T3)} — alternative reasoning path
    \item \texttt{c4: revise\_claim(c2) (T4)} — refined evaluation
    \item \texttt{c5: merge\_claims(c2, c3, c4) (T5)} — integrated final reasoning
\end{itemize}

All claims inherit \texttt{root\_claim\_id = c1}, forming a single coordination cascade.

\paragraph{Coordination Structures.}
Figure~\ref{fig:worked_example_pipeline} shows the full pipeline: 
(left) task expansion into an interdependent structure, 
(center) claim-level DAG constructed from event traces, 
and (right) the cascade extracted from the root claim.

\paragraph{Observable Computation.}
From the cascade rooted at $c_1$, we compute:

\begin{itemize}
    \item \textbf{Revision Wave:} $c_1 \rightarrow c_2 \rightarrow c_4$ (length = 3)
    \item \textbf{Contradiction Burst:} claim $c_1$ receives a competing claim $c_3$ (size = 1)
    \item \textbf{Merge Fan-in:} $c_5$ integrates $c_2, c_3, c_4$ (fan-in = 3)
    \item \textbf{Total Cognitive Effort (TCE):} all reachable claims $\{c_1,\dots,c_5\}$ (size = 5)
\end{itemize}

\paragraph{Interpretation.}
This example reflects the coordination dynamics induced by our workload expansion procedure: 
task decomposition introduces interdependent subtasks, agents generate competing and refined reasoning across these subtasks, and merge operations integrate multiple branches. 
The resulting cascade structure is the fundamental object used to study scaling laws of coordination in our experiments.

\subsection{Observable Trigger Conditions}
\label{app:observable_triggers}

We define the exact extraction rules used to compute coordination observables from reconstructed structures. 
All observables are computed from either the claim DAG, the subtask tree, or root-centered cascades as described in Sections~\ref{app:dag_construction} and~\ref{app:worked_example}.
\begin{table}[t]
\centering
\small
\setlength{\tabcolsep}{5pt}
\renewcommand{\arraystretch}{1.2}

\begin{tabular}{
p{3.2cm}   
p{5.8cm}   
p{2.5cm}   
p{2.2cm}   
}
\toprule
\textbf{Observable} & \textbf{Trigger Condition} & \textbf{Structure} & \textbf{Output} \\
\midrule

Delegation Cascade 
& Subtree rooted at a \texttt{delegate\_subtask} event 
& Subtask tree 
& Node count \\

Revision Wave 
& Maximal chain of claims with identical \texttt{revision\_chain\_id} 
& Claim DAG 
& Length \\

Contradiction Burst 
& Claims referencing the same parent claim within a temporal window $\tau$ 
& Claim DAG + timestamps 
& Count \\

Merge Fan-in 
& Number of \texttt{parent\_claim\_ids} in a \texttt{merge\_claims} event 
& Claim DAG 
& In-degree \\

Total Cognitive Effort (TCE) 
& Reachable subgraph from \texttt{root\_claim\_id} 
& Cascade 
& Size \\

\bottomrule
\end{tabular}

\caption{
Formal trigger conditions for coordination observables. Each observable is computed from structured traces via deterministic extraction rules applied to reconstructed coordination structures.
}
\label{tab:observable_triggers}
\end{table}
\paragraph{Notes.}
All grouping variables (e.g., \texttt{revision\_chain\_id}, \texttt{contradiction\_group\_id}) are derived from raw traces during post-processing and are not directly emitted by the system. 
Temporal windows for contradiction grouping are defined relative to event timestamps. 
Observables are computed consistently across all tasks, topologies, and system scales.

\FloatBarrier
\section{LLM Ablation}
\label{supp-sec:llm-ablation}

\begin{table*}[h]
\centering
\small
\setlength{\tabcolsep}{5pt}
\renewcommand{\arraystretch}{1.12}
\begin{tabular}{lccccccccc}
\toprule
& \multicolumn{3}{c}{\textbf{Tail structure}} & \multicolumn{2}{c}{\textbf{LR tests}} & \textbf{Pref.\ Reinf.} & \multicolumn{2}{c}{\textbf{Concentration}} & \textbf{Outcome} \\
\cmidrule(lr){2-4}
\cmidrule(lr){5-6}
\cmidrule(lr){8-9}
\textbf{Model} & $\hat{\alpha}$ & $\hat{x}_c$ & $x_{\max}$ & $\mathrm{LR}_{\mathrm{T}/\mathrm{LN}}$ & $\mathrm{LR}_{\mathrm{T}/\mathrm{PL}}$ & $\hat{\beta}$ & $E^{\mathrm{all}}_{10}$ & Active & Task Succ. \\
\midrule
GPT-4o-mini         & 2.22 & 57 & 1320 & +5.2 & +2.8 & 0.16 & 25\% & 81\% & 0.48 \\
Qwen 2.5 72B   & 2.27 & 44 & 1035 & +5.0 & +2.7 & 0.15 & 23\% & 79\% & 0.43 \\
Llama 3.1 70B  & 2.34 & 30 & 1012 & +4.8 & +2.5 & 0.13 & 19\% & 78\% & 0.41 \\
Qwen 2.5 7B    & 2.74 & 9  & 440  & +3.5 & +1.7 & 0.07 & 14\% & 72\% & 0.27 \\
\bottomrule
\end{tabular}
\caption{\textbf{Global pooled TCE robustness across models.} 
All 16 topology $\times$ task conditions are pooled per model. For all cross-model TCE fits, we fix the tail onset threshold at $x_{\min}=8$ to maintain a common fitting regime across models. Tail statistics report the fitted truncated-power-law parameters for total collective effort (TCE), together with likelihood-ratio comparisons against log-normal and pure power-law alternatives. Preferential reinforcement is summarized by $\hat{\beta}$, concentration by the top-10\% effort share and active-agent fraction, and outcome by overall task success.}
\label{tab:llm-ablation-global}
\end{table*}

\begin{table*}[t]
\centering
\footnotesize
\setlength{\tabcolsep}{3.8pt}
\renewcommand{\arraystretch}{1.06}
\begin{tabular}{@{}llcccccc@{}}
\toprule
& & \multicolumn{2}{c}{\textbf{Tail structure}} & \multicolumn{2}{c}{\textbf{LR tests}} & \textbf{Concentration} & \textbf{Outcome} \\
\cmidrule(lr){3-4}
\cmidrule(lr){5-6}
\textbf{Topology} & \textbf{Model} & $\hat{\alpha}$ & $\hat{x}_c$ & $\mathrm{LR}_{\mathrm{T}/\mathrm{LN}}$ & $\mathrm{LR}_{\mathrm{T}/\mathrm{PL}}$ & $E^{\mathrm{all}}_{10}$ & Task succ. \\
\midrule
Chain
& GPT-4o-mini        & 2.49 & 23 & +4.6 & +2.6 & 17\% & 0.43 \\
& Qwen 2.5 72B  & 2.52 & 21 & +4.5 & +2.5 & 16\% & 0.40 \\
& Llama 3.1 70B & 2.61 & 18 & +4.2 & +2.3 & 14\% & 0.36 \\
& Qwen 2.5 7B   & 2.88 &  9 & +3.0 & +1.5 & 11\% & 0.19 \\
\midrule
Star
& GPT-4o-mini        & 2.21 & 44 & +5.2 & +2.8 & 32\% & 0.47 \\
& Qwen 2.5 72B  & 2.24 & 41 & +5.1 & +2.7 & 30\% & 0.44 \\
& Llama 3.1 70B & 2.33 & 36 & +4.9 & +2.5 & 27\% & 0.42 \\
& Qwen 2.5 7B   & 2.63 & 11 & +3.6 & +1.7 & 20\% & 0.26 \\
\midrule
Hierarchical
& GPT-4o-mini        & 2.16 & 35 & +5.1 & +2.7 & 21\% & 0.53 \\
& Qwen 2.5 72B  & 2.19 & 32 & +5.0 & +2.6 & 18\% & 0.49 \\
& Llama 3.1 70B & 2.28 & 28 & +4.8 & +2.4 & 15\% & 0.46 \\
& Qwen 2.5 7B   & 2.58 &  9 & +3.4 & +1.6 & 13\% & 0.31 \\
\midrule
Fully Connected
& GPT-4o-mini        & 2.15 & 61 & +5.7 & +3.2 & 32\% & 0.48 \\
& Qwen 2.5 72B  & 2.18 & 56 & +5.6 & +3.1 & 31\% & 0.46 \\
& Llama 3.1 70B & 2.27 & 49 & +5.4 & +2.9 & 28\% & 0.42 \\
& Qwen 2.5 7B   & 2.57 & 16 & +4.1 & +2.1 & 15\% & 0.29 \\
\bottomrule
\end{tabular}
\caption{\textbf{TCE heavy-tail structure by topology and model.}
Results are pooled across task types within each topology. For all cross-model TCE fits, we fix the tail onset threshold at $x_{\min}=8$ to maintain a common fitting regime across models. Tail statistics report the fitted truncated-power-law parameters for total collective effort (TCE), together with likelihood-ratio comparisons against log-normal and pure power-law alternatives. Concentration is summarized by the top-10\% effort share, and outcome by pooled task success.}
\label{tab:llm-ablation-topology}
\end{table*}

\begin{figure*}[h]
    \centering
    \includegraphics[width=0.95\textwidth]{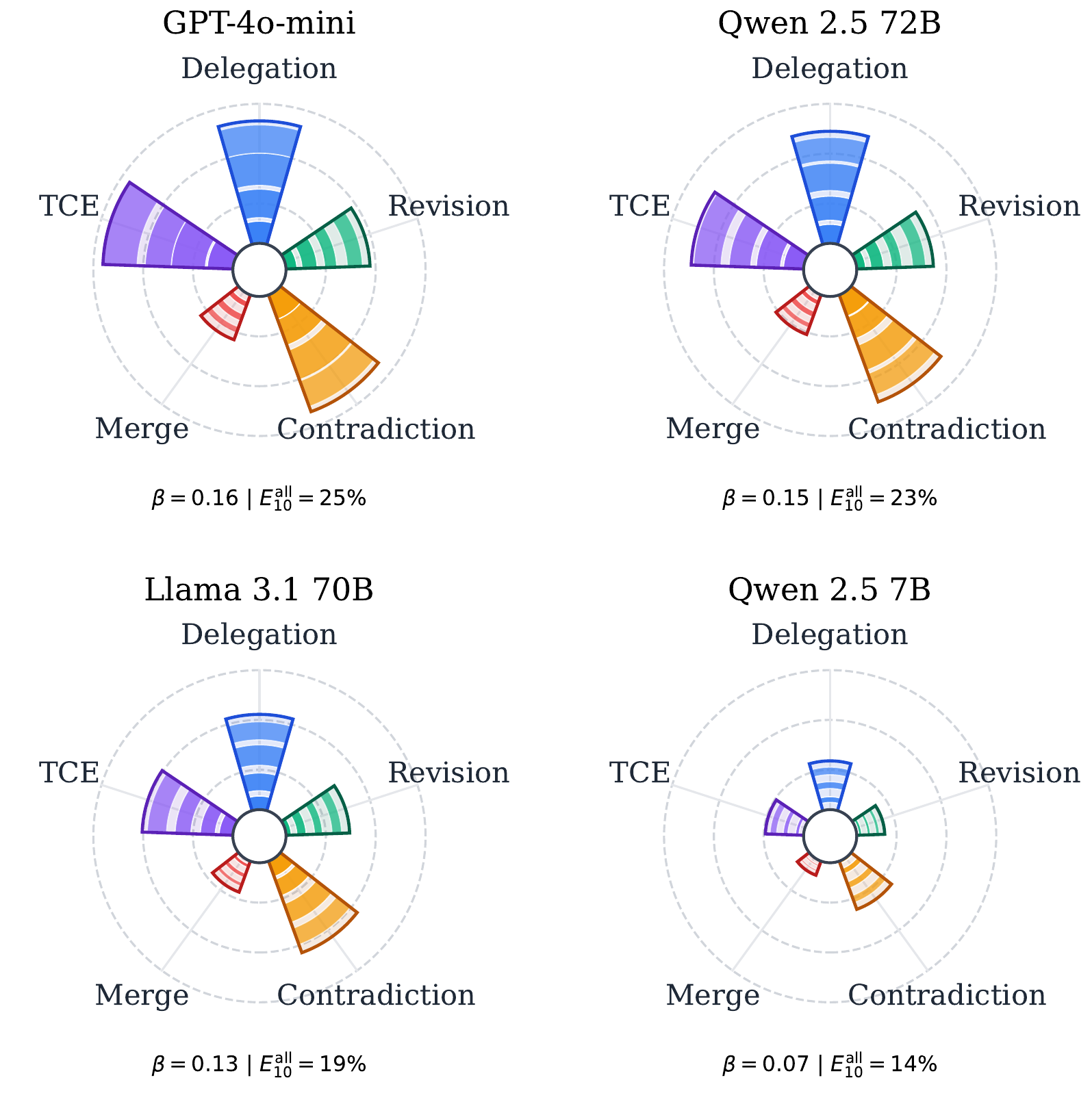}
    \caption{
    \textbf{Coordination-law flower signatures across models.}
    Each panel summarizes one model using five global event observables: delegation cascade, revision wave, contradiction burst, merge fan-in, and total cognitive effort (TCE). Petal extent jointly reflects four law dimensions: heavier tails (lower $\hat{\alpha}$), larger truncation scale $\hat{x}_c$, stronger preferential reinforcement $\hat{\beta}$, and greater elite concentration $E^{\mathrm{all}}_{10}$. Across all four models, the qualitative ordering is preserved: TCE remains the strongest composite coordination signal, delegation and contradiction form the next tier of broad coordination, revision is intermediate, and merge is the weakest and most localized. Moving from GPT-4o-mini to Qwen 2.5 72B, Llama 3.1 70B, and Qwen 2.5 7B, the petals contract inward rather than changing shape abruptly, indicating that heavy-tailed coordination and elite formation persist across model families and scales, but with shorter tails, lower cutoff scales, and weaker concentration in smaller models owing to reduced model-inherent resources.
    }
    \label{fig:llm-ablation}
\end{figure*}

\begin{table*}[h]
\centering
\footnotesize
\setlength{\tabcolsep}{3.6pt}
\renewcommand{\arraystretch}{1.05}
\begin{tabular}{@{}llcccccc@{}}
\toprule
& & \multicolumn{2}{c}{\textbf{Tail structure}} & \multicolumn{2}{c}{\textbf{LR tests}} & \textbf{Concentration} & \textbf{Outcome} \\
\cmidrule(lr){3-4}
\cmidrule(lr){5-6}
\textbf{Task type} & \textbf{Model} & $\hat{\alpha}$ & $\hat{x}_c$ & $\mathrm{LR}_{\mathrm{T}/\mathrm{LN}}$ & $\mathrm{LR}_{\mathrm{T}/\mathrm{PL}}$ & $E^{\mathrm{all}}_{10}$ & Task succ. \\
\midrule
Planning
& GPT-4o-mini        & 2.11 & 54 & +5.2 & +2.2 & 30\% & 0.45 \\
& Qwen 2.5 72B  & 2.14 & 50 & +5.1 & +2.1 & 27\% & 0.41 \\
& Llama 3.1 70B & 2.23 & 44 & +4.8 & +1.9 & 22\% & 0.40 \\
& Qwen 2.5 7B   & 2.53 & 14 & +3.6 & +1.1 & 16\% & 0.23 \\
\midrule
Reasoning
& GPT-4o-mini        & 2.20 & 45 & +5.2 & +2.7 & 23\% & 0.56 \\
& Qwen 2.5 72B  & 2.23 & 42 & +5.0 & +2.6 & 22\% & 0.52 \\
& Llama 3.1 70B & 2.32 & 36 & +4.8 & +2.4 & 19\% & 0.51 \\
& Qwen 2.5 7B   & 2.62 & 11 & +3.6 & +1.6 & 14\% & 0.34 \\
\midrule
Coding
& GPT-4o-mini        & 2.31 & 34 & +5.2 & +3.0 & 24\% & 0.46 \\
& Qwen 2.5 72B  & 2.34 & 31 & +5.0 & +2.9 & 21\% & 0.44 \\
& Llama 3.1 70B & 2.43 & 27 & +4.8 & +2.7 & 18\% & 0.42 \\
& Qwen 2.5 7B   & 2.72 &  9 & +3.5 & +1.9 & 13\% & 0.29 \\
\midrule
QA
& GPT-4o-mini        & 2.40 & 25 & +5.1 & +3.3 & 20\% & 0.51 \\
& Qwen 2.5 72B  & 2.44 & 23 & +5.0 & +3.2 & 19\% & 0.47 \\
& Llama 3.1 70B & 2.52 & 20 & +4.8 & +3.0 & 18\% & 0.44 \\
& Qwen 2.5 7B   & 2.80 &  9 & +3.4 & +2.2 & 12\% & 0.33 \\
\bottomrule
\end{tabular}
\caption{\textbf{TCE heavy-tail structure by task type and model.}
Results are shown for four representative task families and are pooled across topologies, seeds, and agent-society sizes within each task-type $\times$ model condition. For all cross-model TCE fits, we fix the tail onset threshold at $x_{\min}=8$ to maintain a common fitting regime across models. Tail statistics report the fitted truncated-power-law parameters for total collective effort (TCE), together with likelihood-ratio comparisons against log-normal and pure power-law alternatives. Concentration is summarized by the top-10\% effort share $E^{\mathrm{all}}_{10}$, and outcome by pooled task success.}
\label{tab:llm-ablation-tasktype}
\end{table*}

\FloatBarrier
\section{Additional Details on Experimental Setup}
\label{app:exp-setup}

\begin{table}[h]
\centering
\small
\setlength{\tabcolsep}{4pt}
\renewcommand{\arraystretch}{1.1}

\begin{tabular}{p{0.28\linewidth} p{0.68\linewidth}}
\toprule
\textbf{Component} & \textbf{Configuration} \\
\midrule
Benchmarks & GAIA, SWE-bench Verified, REALM-Bench, MultiAgentBench \\
Task types & QA, reasoning, coding, planning \\
Total tasks & $\sim$400 (stratified across benchmarks and difficulty levels) \\
Agent counts ($N$) & $\{8,16,32,64,128,256,512\}$ \\
Topologies & Chain, Star, Tree, Hierarchical, Fully Connected, Sparse Mesh, Dynamic Reputation \\
Agent model & Shared LLM (GPT-4o-mini) \\
Execution steps & 20 per run \\
Seeds & 5 independent random seeds per configuration \\
Task expansion & Benchmark-conditioned expansion module (Appendix ~\ref{app:workload_expansion}) \\
Trace granularity & Event-level (delegation, revision, contradiction, merge, endorsement) \\
\midrule
Total runs & $\approx 400 \times 7 \times 7 \times 5 \approx 98{,}000$ \\
\bottomrule
\end{tabular}
\caption{Experimental configuration summary.}
\end{table}

\begin{table}[h]
\centering
\small
\setlength{\tabcolsep}{4pt}
\renewcommand{\arraystretch}{1.1}

\begin{tabular}{lcc}
\toprule
\textbf{Observable} & \textbf{Samples} & \textbf{Mean/run} \\
\midrule
Delegation cascades & $\sim 3.0 \times 10^5$ & $\sim 3.1$ \\
Revision waves & $\sim 2.8 \times 10^5$ & $\sim 2.9$ \\
Contradiction bursts & $\sim 3.2 \times 10^5$ & $\sim 3.3$ \\
Merge events & $\sim 2.5 \times 10^5$ & $\sim 2.6$ \\
TCE (root cascades) & $\sim 1.7 \times 10^5$ & $\sim 1.7$ \\
\midrule
\textbf{Total} & $\mathbf{> 1.5 \times 10^6}$ & $\mathbf{\sim 15.3}$ \\
\bottomrule
\end{tabular}

\caption{Coordination-event samples extracted from $\sim 98{,}000$ runs. Each run produces multiple event instances depending on the realized interaction trace.}
\end{table}

\FloatBarrier
\section{Workload Expansion Module}
\label{app:workload_expansion}

To study coordination across agent societies of varying size, we introduce a
benchmark-conditioned workload expansion module that scales the number of
tasks with $N$ while preserving task diversity and executability.

The module is explicitly constrained to generate \emph{only workload} and does
not encode or bias any coordination structure. All coordination events
(delegation, revision, contradiction, merge, and total cognitive effort) are
extracted solely from the realized interaction traces during execution.

\begin{figure*}[h]
    \centering
    \includegraphics[width=\textwidth]{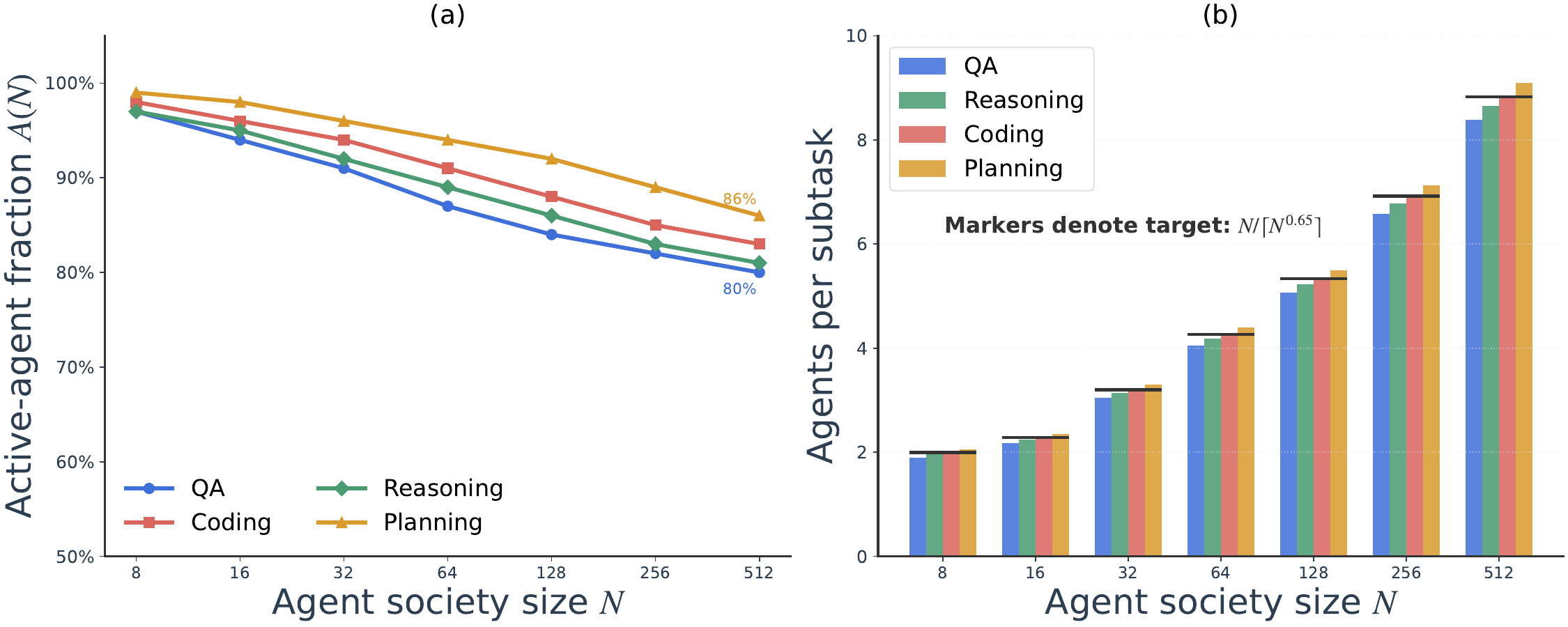}
    \caption{\textbf{Workload-expansion validation across agent society size.}
\textbf{(a)} Active-agent fraction $A(N)$ for four task families remains high across scales,
staying above 80\% even at $N=512$, indicating sustained participation.
\textbf{(b)} Agents per subtask induced by the expansion rule, compared to the target scaling
$N / \lceil N^{0.65} \rceil$. The gradual increase from $\sim$2 to $\sim$9 agents per
subtask shows that workload grows with $N$ without over-concentrating agents.
Together, these results verify that the expansion module maintains balanced workload
and broad agent utilization while not prescribing coordination structure.
}
    \label{fig:task-expansion-appendix}
\end{figure*}

For each benchmark $b$ and task family $d$, we sample $K=\min(5, |B(b,d)|)$ seed
tasks defining a shared problem context. A generator LLM produces related tasks,
which are filtered by a validator using criteria of same-world consistency,
non-paraphrase, independent meaningfulness, executability, and additive
informativeness.

The number of expanded tasks per seed is
\[
M = \left\lceil \frac{N}{K \cdot A} \right\rceil,
\]
with target $A=5$ agents per task, yielding a total workload of $K \times M$
tasks. Tasks are connected via sparse, randomly sampled dependency edges,
forming a shallow DAG independent of the communication topology.

Figure~\ref{fig:task-expansion-appendix} shows that this procedure maintains
high agent utilization ($>80\%$ active at $N=512$) and balanced agents-per-subtask
scaling, confirming that workload grows with $N$ without over-concentration.

\FloatBarrier
\section{Agent Configuration and Experimental Protocol}
\label{app:agent_config}

All agents share a standardized configuration across all topologies, task
families, scales, and seeds. No agent-level hyperparameters are tuned per
condition, ensuring that observed differences arise solely from interaction
structure and workload.

\subsection{Agent Configuration}

\begin{table}[h]
\centering
\small
\setlength{\tabcolsep}{4pt}
\renewcommand{\arraystretch}{1.1}
\begin{tabular}{p{0.35\linewidth} p{0.6\linewidth}}
\toprule
\textbf{Component} & \textbf{Configuration} \\
\midrule
Orchestration & LangGraph (uniform routing, state, and execution) \\
Base prompt & Task, neighbors, structured history  \\
Topology addendum & Topology-specific routing behavior (chain, star, tree, hierarchical, mesh, reputation) \\
Task addendum & Benchmark-specific instructions (QA, coding, planning, reasoning) \\
Tool access & Restricted to benchmark-native tools \\
Context budget & 4000-token window  \\
Completion budget & 1000 tokens \\
Memory & No cross-run persistence; structured state only \\
Seeds & 5 per configuration \\
\bottomrule
\end{tabular}
\caption{Standardized agent configuration used across all experiments.}
\end{table}

\subsection{Benchmarks}

\begin{table}[h]
\centering
\small
\setlength{\tabcolsep}{4pt}
\renewcommand{\arraystretch}{1.1}
\begin{tabular}{lccc}
\toprule
\textbf{Benchmark} & \textbf{Task type} & \textbf{Approx. tasks} & \textbf{Difficulty} \\
\midrule
GAIA & QA / multimodal reasoning & $\sim$150 & Medium–Hard \\
SWE-bench Verified & Code debugging / patching & $\sim$235 & Hard \\
REALM-Bench & Planning / constraint reasoning & $\sim$14 & Hard \\
MultiAgentBench & Coordination / interaction tasks & $\sim$6 scenarios & Variable \\
\bottomrule
\end{tabular}
\caption{Benchmarks used for workload generation and evaluation.}
\end{table}

\subsection{Prompt Structure}

Each agent prompt consists of three layers:
(i) a shared base prompt,
(ii) a topology-specific addendum, and
(iii) a task-family-specific addendum.
The base prompt is identical across all agents and does not prescribe any
coordination strategy.

\paragraph{Base prompt.}
\begin{quote}
You are an AI agent participating in a multi-agent reasoning system.
Your goal is to contribute high-quality reasoning to solve the task.

Core behaviors:
- Think step by step before answering.
- Identify and correct errors in other agents' outputs when necessary.
- Build on useful prior reasoning rather than repeating it.
- Be concise but complete.

Output format:
Provide reasoning followed by a final answer labeled:
\texttt{ANSWER: <your answer>}
\end{quote}

This prompt is intentionally minimal and strategy-agnostic. No explicit
instructions for delegation, critique, or merging are included, ensuring that
coordination patterns emerge from interaction rather than prompt design.

\paragraph{Example topology addendum (reputation routing).}
\begin{quote}
You are agent \{agent\_id\} in a reputation-routed network (step \{step\} of
\{max\_steps\}). You have consulted a set of peers selected by reputation.

- Review consulted outputs critically.
- Do not blindly trust high-reputation agents.
- Revise your reasoning using useful information.
- Prefer correction and synthesis over agreement.

At intermediate steps, produce an improved claim.
At the final step, synthesize the best answer.
\end{quote}

Topology-specific addenda define communication and routing behavior but do not
impose specific coordination patterns. Task-specific addenda provide domain
context (e.g., QA, coding, planning) while maintaining the same interaction
structure.

Overall, this design ensures that coordination dynamics arise from agent
interaction under shared constraints, rather than from prompt engineering or
task-specific scripting.

\end{document}